\documentclass[aps,prb,twocolumn,superscriptaddress,nofootinbib]{revtex4-2}
\usepackage{amssymb,graphicx,color}
\usepackage[intlimits]{amsmath}
\usepackage[english]{babel}
\usepackage[colorlinks]{hyperref}
\hypersetup{
	colorlinks=blue,
	linkcolor=blue,
	filecolor=blue,
	urlcolor=blue,
	citecolor=blue
}
\usepackage[normalem]{ulem}
\usepackage{comment}
\usepackage{ragged2e}
\usepackage{enumerate}
\usepackage{subfigure}
\usepackage{float}
\usepackage{ulem}
\usepackage{soul}

\begin{document}

\title{Periodically and quasiperiodically driven-anisotropic Dicke model }

\author{Pragna Das}
\affiliation{Indian Institute of Science Education and Research Bhopal 462066 India}
\author{Devendra Singh Bhakuni}
\affiliation{Department of Physics, Ben-Gurion University of the Negev, Beer-Sheva 84105, Israel}
\author{Lea F. Santos}
\affiliation{Department of Physics, University of Connecticut, Storrs, Connecticut 06269, USA}
\author{Auditya Sharma}
\email{auditya@iiserb.ac.in}
\affiliation{Indian Institute of Science Education and Research Bhopal 462066 India}

\begin{abstract} 
We analyze the anisotropic Dicke model in the presence of a periodic
drive and under a quasiperiodic drive. The study of drive-induced
phenomena in this experimentally accesible model is important since
although it is simpler than full-fledged many-body quantum systems, it
is still rich enough to exhibit many interesting features. We show
that under a quasiperiodic Fibonacci (Thue-Morse) drive, the system
features a prethermal plateau that increases as an exponential
(stretched exponential) with the driving frequency before heating to
an infinite-temperature state. In contrast, when the model is
periodically driven, the dynamics reaches a plateau that is not
followed by heating. In either case, the plateau value depends on the
energy of the initial state and on the parameters of the undriven
Hamiltonian. Surprisingly, this value does not always approach the
infinite-temperature state monotonically as the frequency of the
periodic drive decreases. We also show how the drive modifies the
quantum critical point and discuss open questions associated with the
analysis of level statistics at intermediate frequencies.
\end{abstract}

\maketitle

\section{Introduction}
\label{sec_1}

The idea of modifying the properties of a system with an external
drive has a long history with early examples including the spin
echo~\cite{Hahn1950} and the Kapitza pendulum~\cite{Kapitza1951}.  The
drive can induce chaos in systems with one degree of freedom, where
chaos is otherwise inaccessible, as in the kicked
rotor~\cite{Casati1979} and the Duffing oscillator~\cite{Ueda1979}.
It can lead to the emergence of double
wells~\cite{Marthaler2006,Zhang2017} that have applications to the
generation of Schr\"odinger cat states~\cite{Puri2017}, and it can
affect the critical point of quantum phase transitions (QPTs) and
excited state quantum phase transitions (ESQPTs)
\cite{caprio2008excited,Cejnar2021}, as verified for the
Lipkin-Meshkov-Glick model~\cite{Chinni2021,Saiz2023}.

In the case of quantum systems with many degrees of freedom, there
have been significant efforts in exploring the use of external drives
to achieve new phases of matter and new physics phenomena not found at
equilibrium. This interest is in part due to experimental advances
that have allowed, for example, the observation of a discrete time
crystal~\cite{Zhang2017crystal}, Floquet prethermalization in dipolar
spin chains~\cite{Peng2021} and in Bose-Hubbard
models~\cite{Rubio2020}, and Floquet topological
insulators~\cite{Pyrialakos2022}. A problem faced by the use of
external drives to engineer Hamiltonians with desired properties is
that the drive usually heats the system to an infinite-temperature
state~\cite{d2014long,Lazarides2014}.  Alternatives that have been
examined to suppress heating involve the inclusion of strong
disorder~\cite{Lazarides2015,Ponte2015}, high-frequency
drive~\cite{Mori2016}, and spectrum
fragmentation~\cite{bhakuni2021suppression}.

In this paper, we focus on the Dicke model~\cite{dicke1954coherence},
which is a many-body system with two degrees of freedom and therefore
bridges the gap between the two extremes mentioned above of systems
with one-degree of freedom and systems with many interacting particles
and many degrees of freedom. We investigate how the Dicke model's
static and dynamical properties change when a periodic external drive
or a quasiperiodic drive is applied. Our analysis addresses
modifications to the quantum critical point, the regular-to-chaos
transition, the onset of a prethermal plateau in the quench dynamics,
how the duration of this plateau depends on the driving frequency, the
energy of the initial state, and the parameters of the undriven
Hamiltonian, and whether the plateau is followed by heating to an
infinite-temperature state.

Introduced as a model of light-matter interaction to explain the
phenomenon of superradiance~\cite{hepp1973superradiant,
  wang1973phase}, the Dicke model describes a system of $N$ two-level
atoms that collectively interact with a single-mode bosonic
field~\cite{dicke1954coherence}. The model can be experimentally
realized with optical
cavities~\cite{baumann2010dicke,baumann2011exploring,arnold2011collective,Baden2014,klinder2015observation,Zhang2018},
trapped ions~\cite{Safavi2018}, and circuit quantum
electrodynamics~\cite{Jaako2016}. Depending on the Hamiltonian
parameters and excitation energies, the undriven system can be in the
regular or chaotic
regime~\cite{emary2003chaos,Chavez2016,buijsman2017nonergodicity}, and
in addition to the normal to superradiant
QPT~\cite{emary2003chaos,emary2003quantum,lambert2004entanglement,zhu2019entanglement,hu2021out},
it also exhibits an
ESQPT~\cite{perez2011excited,Brandes2013,Bastarrachea2014,perez2017thermal,das2021phase,das2023phase,chavez2019quantum,pilatowsky2020positive,Lewis-Swan2019}.
The model has also been used in studies of quantum
scars~\cite{Aguiar1991,Villasenor2020,Pilatowsky2021,Pilatowsky2021NJP},
the onset of the correlation hole (``ramp'')
\cite{lerma2019dynamical}, and thermalization~\cite{Villasenor2023}.

Under a periodic drive, the analysis of the Dicke model has 
focused on the normal to the super-radiant phase and 
chaos~\cite{bastidas2012nonequilibrium,dasgupta2015phase,ray2016quantum}.
We extend these studies to the anisotropic Dicke
model~\cite{hioe1973phase,kloc2017quantum,buijsman2017nonergodicity,aedo2018analog,shapiro2020universal,hu2021out,das2023phase},
which is a generalization to the case of two independent light-matter
couplings. This version of the model is also experimentally
accessible~\cite{Zou2014}. We show that the normal phase is stretched
under a high-frequency periodic drive and, using the Magnus
expansion~\cite{Blanes2009}, we establish a modified condition for the
normal to the superradiant transition.

For the periodically driven system, we also investigate level
statistics and find that at intermediate frequencies, the results
suggest regularity even when the undriven system is chaotic. In
contrast, the evolution of the average boson
number~\cite{emary2003chaos} and of the entanglement
entropy~\cite{lambert2004entanglement,bhakuni2018characteristic}
indicate a degree of spreading in the Hilbert space that is at least
equivalent to that reached by the undriven system, which implies that
the results for level statistics may be an artifact. An intriguing
element to this picture is that for high-energy initial states, there
is a narrow range of intermediate frequencies for which the saturation
value of the average boson number becomes larger than the
infinite-temperature result. We believe that this is caused by a lack
of full ergodicity and that near equipartition only happens for small
driving frequencies.

The core of this paper is the comparison of the dynamics of the
anisotropic Dicke model under periodic and quasiperiodic drives, which
show distinct behaviors. When periodically driven, it saturates to a
plateau that is not followed by heating to the infinite-temperature
state. The saturation value depends on the frequency of the drive, the
energy of the initial state, and whether the undriven system is in the
regular or chaotic regime. The spreading of low-energy initial states
at intermediate to high frequencies is very restrained. In contrast,
under a quasiperiodic drive modeled by the
Thue-Morse~\cite{thue1906uber,nandy2017aperiodically,mukherjee2020restoring,
  zhao2021random,mori2021rigorous,tiwari2023dynamical}
(Fibonacci~\cite{dumitrescu2018logarithmically,tiwari2023dynamical,maity2019fibonacci,
  ray2019dynamics}) sequence, the model presents a prethermal plateau
that grows as a stretched exponential (exponential) with the driving
frequency and is later followed by heating. This is similar to what
was found for many-body spin models, where the heating time was shown
to grow exponentially with the driving frequency for the Fibonacci
drive protocol~\cite{dumitrescu2018logarithmically}. In contrast,
under the Thue-Morse protocol, it was found~\cite{mori2021rigorous}
that the heating time is shorter than exponential and longer than
algebraic in the driving frequency.

The presence (absence) of the heating process for quasiperiodic
(periodic) drives is aligned with the discussion
in~\cite{Pilatowsky2023}, where complete Hilbert-space ergodicity was
proven for systems under nonperiodic drives, but discarded for
time-independent or time-periodic Hamiltonian dynamics. Paradoxically,
there are results that indicate prethermalization followed by heating
in periodically driven many-body spin systems with short- and
long-range interactions~\cite{machado2019exponentially} and in
periodically driven arrays of coupled kicked rotors~\cite{Rajak2019},
although it might be that these systems do not reach full ergodicity
in the sense presented in~\cite{Pilatowsky2023}.

\section{Model Hamiltonian}
\label{sec_2}

The Hamiltonian of the generalized Dicke model with time-dependent couplings is given by
\begin{eqnarray}
{\cal H}(t) &=& \omega a^{\dagger}a + \omega_{0}J_{z} + \frac{\tilde{g}_{1}(t)}
{\sqrt{2j}}(a^{\dagger}J_{-} + a J_{+})\nonumber\\
&& + \frac{\tilde{g}_{2}(t)}{\sqrt{2j}}(a^{\dagger}J_{+} + a J_{-}) ,
\label{eqn:hamiltonian}
\end{eqnarray} 
where we have set $\hbar=1$; $a$ and $a^{\dagger}$ are the creation
and annihilation bosonic operators with $[a,a^{\dagger}]=1$;
$J_{\pm,z}=\sum_ {i=1}^{2j}\frac{1}{2}\sigma_{\pm,z}^{(i)}$ represent
the angular momentum operators of a pseudospin consisting of $N=2j$
two-level atoms described by Pauli matrices $\sigma_{\pm,z}^{(i)}$,
which act on site $i$ and satisfy the relations $[J_z,J_{\pm}]=\pm
J_{\pm}$, $[J_+,J_-]=2J_z$; $\omega$ is the mode frequency of the
bosonic field; $\omega_0$ is the level splitting of the atoms; the
parameters $\tilde{g}_{1}(t)$ and $\tilde{g}_{2}(t)$ are,
respectively, the time-dependent rotating and counter-rotating
interaction terms of the light-matter coupling. For all of our
numerical results, we fix $\omega, \omega_0 =1$.

The Hilbert space is spanned by the basis states $\vert {\cal
  B}_{n,m}\rangle = \{\vert n\rangle\otimes \vert j,m\rangle\}$, where
$\vert n\rangle$ are the Fock states, $a^{\dagger}a\vert n\rangle=
n\vert n\rangle$, and $\vert j,m\rangle$ are the eigenstates of
$J_{\pm,z}$ with $J_{\pm}\vert j,m\rangle=\sqrt{j(j+1)-m(m\pm 1)}
\vert j,m \pm 1\rangle$. To perform our numerical calculations, the
Hilbert space of the bosonic modes is truncated to a finite number
$n_{\max}$, which is large enough to guarantee convergence, that is,
by increasing $n_{\max}$ one does not see qualitative changes in the
calculated quantities. The total truncated Hilbert space dimension is
${\cal N}=(N + 1)(n_{\text{max}} + 1)$.

The finite undriven system presents a precursor of a second-order QPT
from the normal to the superradiant phase~\cite{das2021phase}, which
takes place in the thermodynamic limit ($N\to \infty$), and presents a
transition from the regular to the chaotic
regime~\cite{buijsman2017nonergodicity} that depends on the coupling
parameters and the excitation energies. The point for the two
transitions do not necessarily coincide. In the absence of the
counter-rotating term, when $\tilde{g}_{2}(t)=0$ and
$\tilde{g}_{1}(t)=g_1$, Hamiltonian (\ref{eqn:hamiltonian}) describes
the Tavis-Cummings model, which is regular for any excitation energy.

The undriven Dicke model has two degrees of freedom.  In systems with
few degrees of freedom and a properly defined classical limit, such as
the Dicke model, the notion of quantum chaos is well established. It
refers to properties of the spectrum -- level repulsion and rigidity,
in particular -- that signal chaos in the classical limit, where the
Lyapunov exponent is positive and there is mixing. A parallel between
the values of the Lyapunov exponent and the degree of level repulsion
for the Dicke model with $g_1=g_2$ can be found
in~\cite{Villasenor2023}, where it is seen that classical and quantum
chaos are evident for strong interaction and large excitation
energies. In the present paper, we use the terms ``quantum chaos'' and
``quantum ergodicity'' as synonyms.

In what follows, we investigate how the properties of the generalized
Dicke model change under a time-dependent periodic drive
[Sec.~\ref{sec_3PERIODIC}] and under a quasiperiodic drive
[Sec.~\ref{sec_3APERIODIC}].

\section{Periodic drive}
\label{sec_3PERIODIC} 
The periodic driving protocol that we consider is
\begin{align}
\tilde{g}_i(t) = & 
g_i + \Omega \ \text{Sgn}(\sin \omega_{\text{d}} t),
\end{align}
where $i=1,2$ identifies the two coupling parameters, $g_i$ are positive constants, $\Omega$ is the amplitude of the drive, Sgn[.] is the sign function, and $\omega_{\text{d}} = 2\pi/T$ is the frequency of the drive. The unitary operator over a cycle is constructed as 
\begin{equation}
U(T)=e^{-iH_{B}T/2} e^{-iH_{A}T/2}\equiv e^{-iH_{F}T},
\end{equation}
where 
\begin{equation}
H_{A} = H + V \hspace{0.5 cm } \text{and} \hspace{0.5 cm } H_{B} = H - V,
\end{equation}
$$
H = \omega a^{\dagger}a + \omega_{0}J_{z}  
+ \frac{g_1 }
{\sqrt{2j}}(a^{\dagger}J_{-} + a J_{+})  + \frac{g_2}{\sqrt{2j}}(a^{\dagger}J_{+} + a J_{-}),  
$$
$$
V =
 \frac{\Omega}
{\sqrt{2j}}(a^{\dagger}J_{-} + a J_{+})  + \frac{ \Omega}{\sqrt{2j}}(a^{\dagger}J_{+} + a J_{-}), 
$$
and $H_F$ is the time-independent Floquet Hamiltonian. The unitary operator can be decomposed as $U(T)=\sum_{\nu}e^{-i\phi_{\nu}}\vert\varphi_{\nu}\rangle \langle\varphi_{\nu}\vert$, where $\phi_{\nu}$ are the Floquet phases and $\epsilon_{\nu} =  \mod[\phi_{\nu},2\pi]/T$ are the quasienergies, and $\vert\varphi_{\nu}\rangle$ are the corresponding Floquet modes~\cite{Holthaus2016}.

\subsection{Quantum Phase Transition}

\begin{figure}[h]
	\includegraphics[scale=0.54]{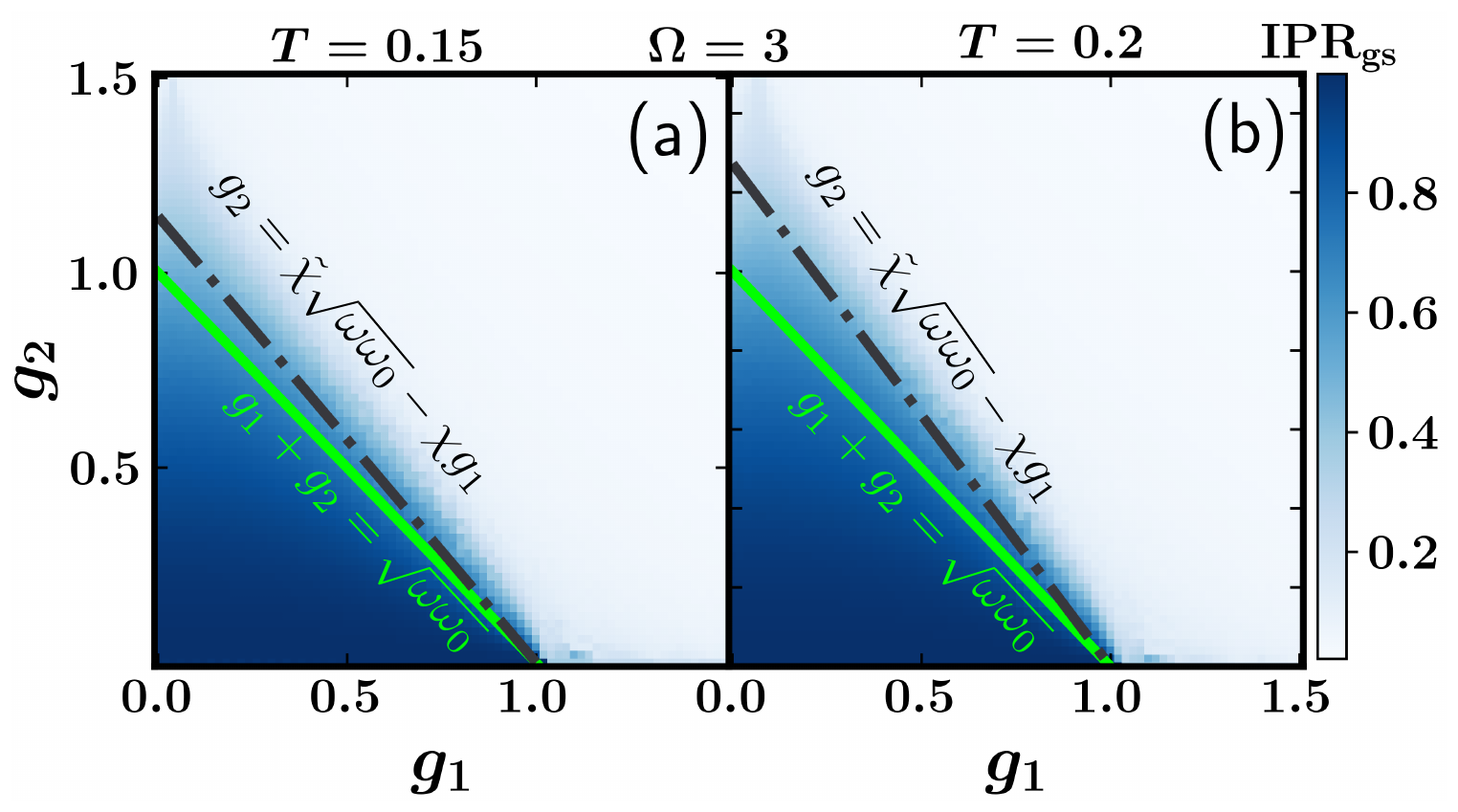}
	\caption{Inverse participation ratio of the ground state of the effective Hamiltonian in Eq.~(\ref{eqn:effective_hamiltonian}) as a function of the coupling parameters $g_1$ and $g_2$;  $\Omega=3$ and $T$ is indicated. The green solid line represents the critical line for the QPT of the undriven case and the black dashed-dotted line is the modified critical line for the periodically driven system;  $N=10$ and $n_{\text{max}}=199$.}
	\label{fig:IPR}
\end{figure}

We start our analysis with a discussion of how the quantum critical
point depends on the drive.  The critical point for the undriven
system is given by $g_1 + g_2 =
\sqrt{\omega\omega_0}$~\cite{buijsman2017nonergodicity} and is marked
with a green solid line in Fig.~\ref{fig:IPR}. To see how this gets
modified by the periodic drive, we perform the Magnus expansion and
obtain an effective Hamiltonian $H_{\text{eff}}$ up to second order in
$\omega_{\text{d}}$ (see details in appendix~\ref{app_1a}):
\begin{eqnarray}
H_{\text{eff}} &=&\omega a^{\dagger}a + \omega_{0}J_{z} + \frac{g_{1}}{\sqrt{2j}}(a^{\dagger}J_{-} + a J_{+})\nonumber\\ 
&+&  \frac{g_{2}}{\sqrt{2j}}(a^{\dagger}J_{+} + a J_{-}) - \frac{T^{2}}{12}\Big[-\frac{4\omega\Omega^{2}}{N}J_x^{2}\nonumber\\ 
&+&  \frac{2\omega_0 \Omega^{2}}{N}(a^{\dagger}+a)^2 J_z + \frac{(g_1 - g_2 )\Omega^{2}}{N\sqrt{N}}\Big(8(a^{\dagger}+a)J_xJ_z \nonumber\\
&+&  (a^{\dagger} - a)({a^{\dagger} + a})^{2}(J_+ - J_-)\Big)\Big].  
\label{eqn:effective_hamiltonian}
\end{eqnarray}
Taking the limit $N\rightarrow \infty$ (see appendix~\ref{app_1}), we
arrive at a modified condition for the normal to the superradiant
transition that holds for $T^2 \Omega^2 <1$ and depends on the period
and amplitude of the drive as
\begin{equation} 
g_2 \approx \tilde{\chi}\sqrt{\omega\omega_0} - \chi g_1  ,
\label{eqn:qpt_line}
\end{equation} 
where
$$ \chi = \frac{1+\delta}{1-\delta}, \hspace {0.5 cm}  \tilde{\chi} = \frac{1+\tilde{\delta}}{1-\delta} $$ 
and 
$$\delta = \frac{T^2\Omega^2}{3},  \hspace {0.5 cm}  \tilde{\delta} = \frac{\delta}{2}\left( \frac{\omega}{\omega_0} + \frac{\omega_0}{\omega} \right).$$

The line determined by Eq.~(\ref{eqn:qpt_line}) is marked with a black
dashed-dotted curve in Fig.~\ref{fig:IPR}. In comparison with the
green line for the undriven system, one sees that with proper choices
of the driving parameters $T$ and $\Omega$, the normal phase can be
extended. Fig.~\ref{fig:IPR} corresponds to the ground-state phase
diagram for the effective Hamiltonian $H_\text{eff}$ in
Eq.~(\ref{eqn:effective_hamiltonian}). The different shades of blue
indicate the numerical value of the inverse participation ratio
$$
\text{IPR}_{\text{gs}} = \sum_{n,m} |c_{n,m}|^4,
$$ of the ground state $|\psi_{\text{gs}}\rangle$, where $c_{n,m} =
\langle {\cal B}_{n,m}|\psi_{\text{gs}}\rangle$.  This quantity
measures the level of delocalization of the ground state with respect
to the basis states. When the ground state coincides with a basis
state, $\text{IPR}_{\text{gs}} =1 $, while a very delocalized ground
state has $\text{IPR}_{\text{gs}} \propto {\cal N}^{-1} $. In
Fig.~\ref{fig:IPR}, darker tones of blue indicate more
localization. The values of $\text{IPR}_{\text{gs}}$ are shown as a
function of the coupling parameters $g_1$ and $g_2$ for two values of
the driving period, namely $T=0.15$ [Fig.~\ref{fig:IPR}(a)] and $T=
0.2$ [Fig.~\ref{fig:IPR}(b)].  The abrupt separation between dark blue
(normal phase) and light blue (superradiant phase) coincides with the
critical line (black dashed-dotted line) obtained in
Eq.~(\ref{eqn:qpt_line}). The panels make it clear that as the period
increases ($\omega_{\text{d}}$ decreases), the critical line appears at larger
values of the coupling parameters, which indicates that the normal
phase gets extended.

\subsection{Level Statistics}

As mentioned above, the anisotropic Dicke model presents regular and
chaotic regimes that can be identified in the quantum domain with the
analysis of level statistics. Here, we investigate how the two regimes
get affected by the presence of the periodic drive. For this, we
consider the ratio of consecutive levels, defined
as~\cite{d2014long,atas2013distribution}
$$r_{\nu} =
\frac{\text{min}(s_{\nu - 1}, s_{\nu})}{\text{max}(s_{\nu - 1},s_{\nu})},
$$ where $s_{\nu}=\epsilon_{\nu+1}-\epsilon_{\nu}$ is the spacing
between consecutive quasienergies (or between consecutive eigenvalues
in the case of time-independent Hamiltonians). In the regular regime,
where the nearest neighboring level spacing distribution is
Poissonian, the average level spacing ratio $\langle r\rangle\approx
0.386$. For chaotic systems described by time-dependent Hamiltonians,
level statistics depends on the driving frequency. If the frequency is
high and ${\cal H}(t)$ is well described by a chaotic static effective
Hamiltonian that is real and symmetric, thus exhibiting time-reversal
symmetry, the level spacing distribution follows the Gaussian
orthogonal ensemble (GOE) and $\langle r\rangle\approx 0.536$. On the
other hand, if the frequency is small and $U(T)$ is a symmetric
unitary matrix, level statistics follows that of a circular orthogonal
ensemble (COE) and $\langle r\rangle\approx 0.527$
\cite{d2014long}. In finite systems, the repulsion is slightly
stronger for GOE than for COE, but the results for both ensembles
should coincide in the thermodynamic limit~\cite{d2014long}.

\begin{figure}[h]
  \includegraphics[scale=0.28]{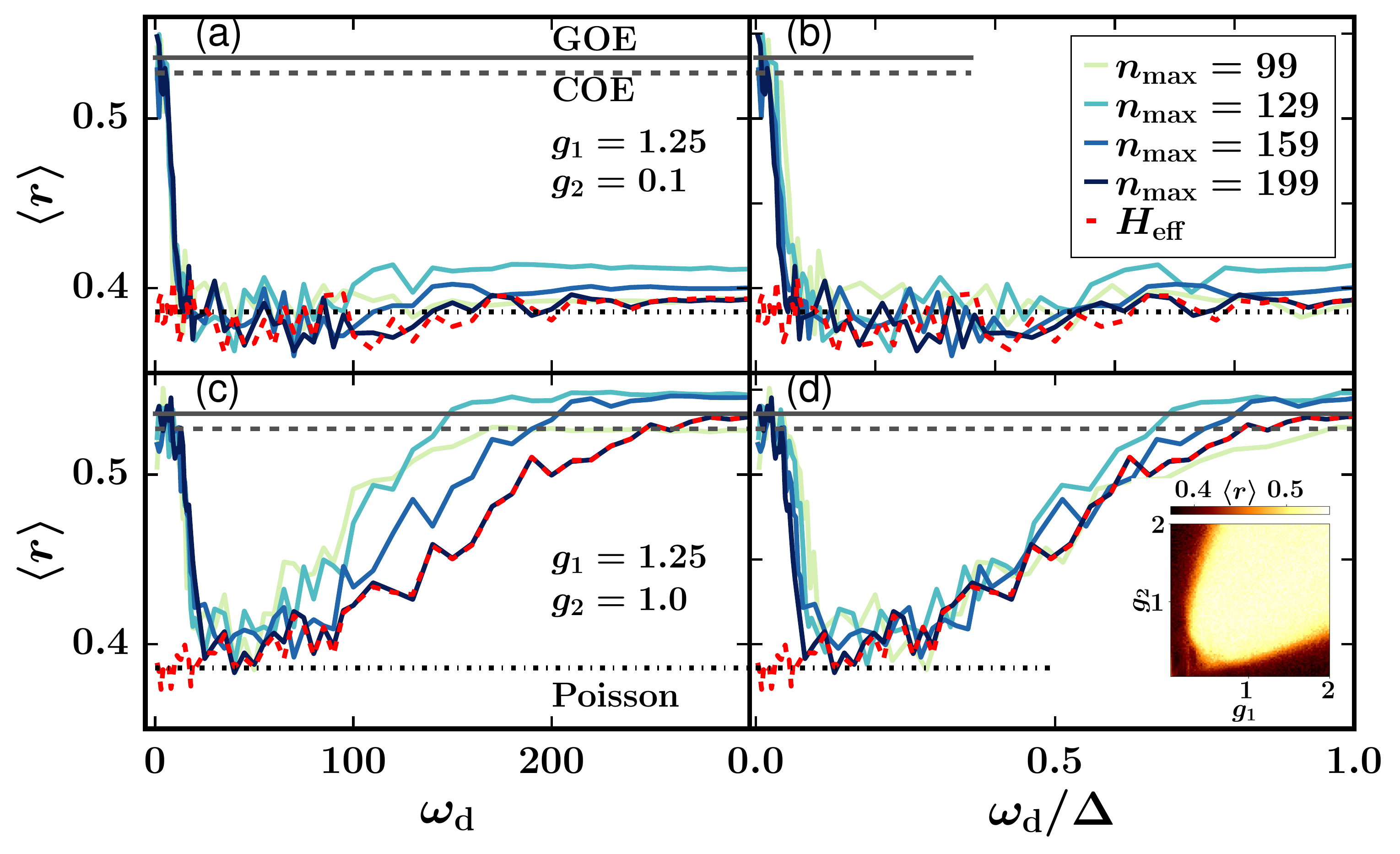}
  \caption{Average consecutive level spacing ratio, $\langle r \rangle$, of the anisotropic Dicke model as a
    function of the driving frequency (a,c) and of the driving 
    frequency rescaled by the energy bandwith of the undriven system  (b,d)
    for $g_2=0.1$ (a)-(b) and $g_2=1.0$  (c)-(d). We fix $g_1=1.25$, $\Omega=1$, and  
    atom number $N=10$. Different bosonic cut-offs are used as indicated in the legend. The inset of panel (d) 
    gives $\langle r \rangle$ as a function of $g_1$ and $g_2$ for the 
    undriven model. 
  }
  \label{fig:ravg}
\end{figure}

For the undriven anisotropic Dicke model, chaos emerges for large
values of the coupling parameters $g_1$ and $g_2$, as shown in the
inset of Fig.~\ref{fig:ravg}(d). Lower and upper band energies, which
are in the nonchaotic region, are discarded for the analysis of level
statistics. We use this figure as a reference for our choices of $g_1$
and $g_2$ in the driven scenario.  The main panels in
Fig.~\ref{fig:ravg} display the average level spacing ratio for the
driven system using different values of the bosonic cutoff
$n_{\text{max}}$. The results are shown as a function of the driving
frequency in Figs.~\ref{fig:ravg}(a,c) and as a function of the
driving frequency rescaled by the energy bandwidth $\Delta$ of the
undriven system in Figs.~\ref{fig:ravg}(b,d). The purpose of the
rescaling is to check the convergence of the results. The solid lines
for the different values of $n_{\text{max}}$ in
Figs.~\ref{fig:ravg}(b,d) are indeed close and, for large frequencies,
they nearly coincide with the curve for the effective Hamiltonian from
Eq.~(\ref{eqn:effective_hamiltonian}) (dashed line), the agreement
being excellent for the largest value of $n_{\text{max}}=199$. Notice
that our $H_{\text{eff}}$ depends on the value of $T$, which contrasts
with similar plots from previous studies, where the effective
Hamiltonian used was obtained to zeroth-order of the Magnus
expansion~\cite{d2014long}.

For the chosen coupling parameters in Figs.~\ref{fig:ravg}(a,b), the
undriven system is regular, while in Figs.~\ref{fig:ravg}(c,d) it is
chaotic. This explains why, at high frequencies, $\langle r \rangle$
in Figs.~\ref{fig:ravg}(a,b) reaches Poisson values, while $\langle r
\rangle$ in Figs.~\ref{fig:ravg}(c,d) reaches GOE values. The
saturation at the GOE value for large $\omega_{\text{d}}$ is more evident for
the largest $n_{\text{max}}$. At low frequencies, the effective
Hamiltonian ceases to be valid and the system becomes chaotic,
independently of the regime of the undriven case. In this case,
$\langle r \rangle$ should approach the COE value.

This last paragraph is dedicated to a possible explanation of what
happens at the intermediate frequencies in Figs.~\ref{fig:ravg}(c,d),
where one sees a significant dip in the values of $\langle r
\rangle$. This may not be caused by a transition to a regular regime
and may instead be an artifact of the process of folding the
quasienergies to the principal Floquet zone $\left[-\omega_{\text{d}}/2,
  \omega_{\text{d}}/2\right]$.  We discuss why we suspect this might be the
case, but a final answer requires the analysis of the system in the
classical limit~\cite{das2023classical}. As noticed
in~\cite{d2014long} and clearly explained
in~\cite{regnault2016floquet}, at intermediate frequencies, some of
the quasienergies lie outside the principal Floquet zone and need to
be folded back. In this process, the folded quasienergies may not
repel the energies originally inside the zone, resulting in a reduced
value of $\langle r \rangle$. This contrasts with the case of a
driving frequency larger than the many-body bandwidth $(\omega_{\text{d}} \gg
\Delta)$, where the reconstruction of the spectrum of quasienergies is
not required and the picture is analogous to that of a
time-independent GOE Hamiltonian. It also contrasts with the case of
low frequency, where the majority of the quasienergies need to be
folded back and one reaches the scenario of COE statistics. It calls
attention, however, that instead of a small dip suggesting a mixed
scenario with some levels still repelling each other, as seen
in~\cite{d2014long, regnault2016floquet}, our results for $\langle r
\rangle$ reach Poisson values and the dip does not diminish as
$n_{\text{max}}$ increases. We blame this result to the strong
asymmetric shape of the density of states. It may be that at
intermediate frequencies, the folded levels affect the states at high
excitation energies, for which the GOE statistics used to hold, while
the states at lower energies, which are not chaotic, do not get
affected. Our speculation finds support in the quantum dynamics
described in the next subsection, where despite the Poisson values
associated with $\langle r \rangle$, the quantum evolution suggests
spreading of the initial state at least comparable to what happens to
the chaotic undriven Hamiltonian. However, we call attention to the
puzzling results in Fig.~\ref{fig:Fig03new}(b) and
Fig.~\ref{fig:Nav_wd}.

\subsection{Dynamics and Dependence on the Initial State}

To study the dynamics, we consider the average boson number, defined as
\begin{equation}
N_\text{av} (t) = \langle \Psi(t)|a^{\dagger}a|\Psi(t)\rangle ,
\label{Eq:Nav}
\end{equation}
where $|\Psi(0)\rangle$ is the initial state, and the von Neumann entanglement entropy between the spins and bosons:
\begin{equation}
S(t) =
-\text{Tr}\left[\rho_{\text{spins}}(t)
\ln(\rho_{\text{spins}}(t))\right],
\label{Eq:S}
\end{equation}
where $\rho_{\text{spins}} (t) = \text{Tr}_{\text{bosons}}\left[\rho
  (t) \right]$ is the reduced density matrix of the spins obtained by
tracing over the bosonic degrees of freedom.

One expects generic driven systems to heat up and reach an
infinite-temperature-like state with
$\rho^{\infty}=\mathcal{I}/\mathcal{N}$ where $\mathcal{I}$ is the
identity matrix and $\mathcal{N}$ is the Hilbert-space dimension. The
infinite-temperature value of the average boson number for the Dicke
model corresponds to
$$N_\text{av}^{\infty}=\text{Tr}\left[\rho^{\infty}_{\text{bosons}}
  a^\dagger a\right] = n_{\text{max}}/2,$$ where
$\rho_{\text{bosons}}^{\infty} =
\text{Tr}_{\text{spins}}\left[\rho^{\infty}\right]$, and the
entanglement entropy saturates to the Page
value~\cite{page1993average}, given by
$$S_{\text{Page}} = \text{ln}(N+1) - \frac{N+1}{2(n_{\text{max}} + 1)}.$$ 

The Page value is derived for bounded systems, while the Hilbert space
of the bosonic subspace of the Dicke model is unbounded. Yet, the
truncation to $n_{\text{max}}$ still provides a meaningful result for
the converged states.  In what follows, we fix the atom number to
$N=10$ and the bosonic mode cut-off at $n_{\text{max}}=199$, which
gives $N_{\text{av}}^{\infty}\approx 100$ and $S_{\text{page}} \approx
2.37$. Our initial states are eigenstates of the decoupled Hamiltonian
($\tilde{g}(t)=0$). We average the data over 50 initial states.

\begin{figure}[h]
	\includegraphics[scale=0.26]{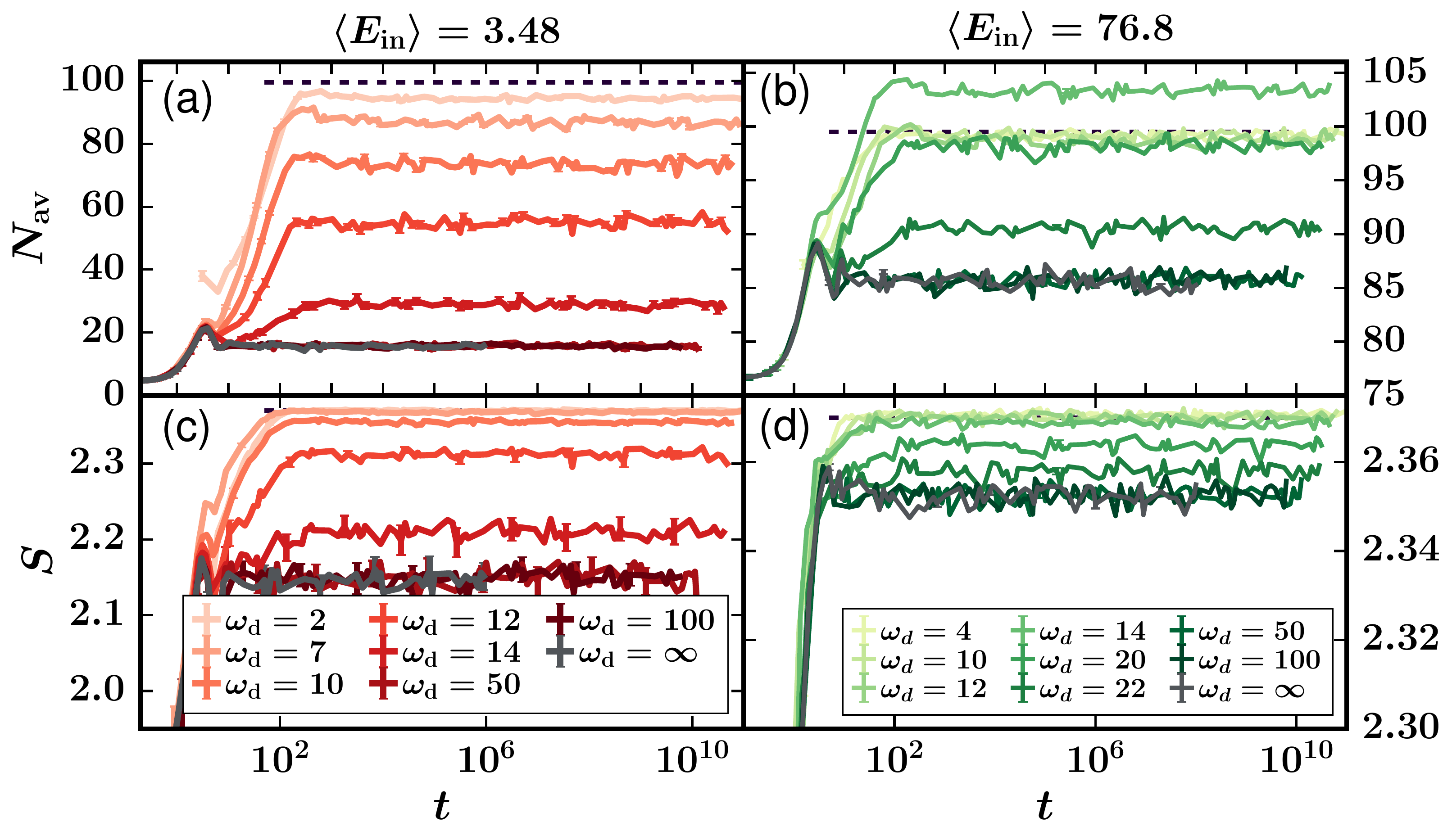}
	\caption{(a,b) Average boson number, $N_\text{av}(t)$, and (c,d) von-Neumann entanglement entropy $S(t)$ as a function of the stroboscopic time $t_n=n T$ for the periodically driven anisotropic Dicke model. Panels (a,c): low energy initial states, so that $\langle E_{\text{in}}\rangle = 3.48$. Panels (b,d): high energy initial states, so that $\langle E_{\text{in}}\rangle = 76.8$. The parameters $g_1 = 1.25,\ g_2 = 1.0$ guarantee chaos in the absence of a drive. The driving frequencies are indicated; the gray line represents the evolution under the effective time-independent Hamiltonian in Eq.~(\ref{eqn:effective_hamiltonian}); the black dashed indicates the results for the infinite-temperature state.  For all plots the driving amplitude $\Omega=1.0$, $N=10$, $n_{\text{max}}=199$. 
	}
	\label{fig:Fig03new}
\end{figure}

In Fig.~\ref{fig:Fig03new}, we select coupling parameters
corresponding to the chaotic undriven model and analyze the evolution
of $N_\text{av} (t)$ [Figs.~\ref{fig:Fig03new}(a)-(b)] and $S(t)$
[Figs.~\ref{fig:Fig03new}(c)-(d)] for initial states with low
[Figs.~\ref{fig:Fig03new}(a,c)] and high
[Figs.~\ref{fig:Fig03new}(b,d)] energies, and under various choices of
the driving frequency. The results are compared with the dynamics for
the time-independent effective Hamiltonian in
Eq.~(\ref{eqn:effective_hamiltonian}) (indicated as $\omega_{\text{d}} =
\infty$ in the figure) and with the result for the
infinite-temperature state (black dashed line).

In Figs.~\ref{fig:Fig03new}(a,c), where the initial states have low
energy, as the driving frequency decreases and level statistics moves
from GOE to COE, the saturation values for $N_\text{av} (t)$ and
$S(t)$ increase monotonically, going from agreement with the result
for the chaotic effective Hamiltonian to agreement with the
infinite-temperature state. Nothing in the figure suggests any special
feature for intermediate frequencies that would justify associating
the dip for $\langle r \rangle$ seen in Fig.~\ref{fig:ravg} with an
enhancement of regular behavior. Below, after some additional
discussions about the low-energy initial states, we investigate what
happens when the initial states have high energies. In this case, a
non-monotonic behavior with $\omega_{\text{d}}$ emerges, but only for the
saturation values for $N_\text{av} (t)$ and in a very narrow range of
intermediate values of the driving frequency.

For high and intermediate driving frequencies, where $H_{\text{eff}}$
approximately describes the system, the saturation values for
$N_\text{av} (t)$ and $S(t)$ found in Figs.~\ref{fig:Fig03new}(a,c)
decrease if we decrease the value of $g_2$ (see Fig.~\ref{fig:FigApp2}
in appendix~\ref{app_2}). This is expected, because decreasing $g_2$
brings the effective Hamiltonian closer to the regular regime.  The
limited spread in the Hilbert space of the low energy states seen in
Figs.~\ref{fig:Fig03new}(a,c), despite the drive and the chaoticity of
$H_{\text{eff}}$, evokes the discussions in
Ref.~\cite{fleckenstein2021prethermalization}, where long-lived
prethermal plateaus were observed for driven many-body spin chains
under periodic drives at intermediate frequencies. It is possible that
the spectrum of our model at low energies presents some special
feature, such as a commensurate structure, that the periodic drive
with intermediate frequencies cannot overcome. This is a point that
deserves further investigation.

Under the periodic drive, one can increase the saturation values of
the average boson number and the entanglement entropy by increasing
the energies of the initial states, as seen in
Figs.~\ref{fig:Fig03new}(b,d). Notice that the scale in the $y$-axis
of these panels is not the same as in
Figs.~\ref{fig:Fig03new}(a,c). For high-energy initial states, as seen
in Fig.~\ref{fig:Fig03new}(d), the saturation values of $S(t)$ become
close to the infinite-temperature state not only for low frequencies,
but also for a range of intermediate frequencies. The results for the
average boson number are, however, intriguing. Contrary to what we see
for the entropy, the saturation value of $N_\text{av}(t)$ does not
increase monotonically to the infinite-temperature result as we
decrease $\omega_{\text{d}}$. Instead, for $\omega_{\text{d}} \lesssim 20$, we observe
that $N_\text{av}^{\text{sat}} > N_\text{av}^{\infty}$ (see results
for $N_\text{av}^{\text{sat}}$ vs $\omega_{\text{d}}$ and for $S^{\text{sat}}$
vs $\omega_{\text{d}}$ for different values of the initial state energy in
Fig.~\ref{fig:Nav_wd} of the appendix~\ref{app_2}). The overshooting
suggests lack of equipartition and predominant contributions from
states with large average boson number. This means that for all
driving frequencies $\omega_{\text{d}} \gtrsim 5$, even when
$N_\text{av}^\text{sat}$ crosses $N_\text{av}^\infty$, there is no
ergodicity, as supported by the saturating values of the entropy,
which for this range of driving frequencies give
$S^\text{sat}<S^\infty$.

The results in Fig.~\ref{fig:Fig03new} and Fig.~\ref{fig:FigApp2} are
in stark contrast to what we observe for the quasiperiodic drive,
where after a transient time, heating does take place. As we show in
the next section, even for intermediate to high frequencies and small
$g_2$, the quasiperiodic drive is capable of bringing the system to
the infinite-temperature state after prethermalization. In
Fig.~\ref{fig:Fig03new}, no matter how far in time we went, we never
saw $N_\text{av} (t)$ and $S(t)$ getting away from their plateaus
towards the infinite-temperature results. The periodically driven
Dicke model with intermediate to high frequencies is thus well
protected against heating, specially when it is prepared in a
low-energy state.

\section{Quasiperiodic drive}
\label{sec_3APERIODIC}

We now consider the case where the time-dependent drive is
quasiperiodic, consisting either of Thue-Morse or Fibonacci
sequences. The Thue-Morse
sequence~\cite{thue1906uber,nandy2017aperiodically,mukherjee2020restoring,zhao2021random,mori2021rigorous,tiwari2023dynamical}
is constructed with unitary operators $U_{\pm}=\exp(-iH_{\pm}T)$, so
that it starts with $U_1 = U_-U_+$ and is followed by $\tilde{U}_1 =
U_+U_-$. Next, $U_2 = \tilde{U}_1 U_1$ is followed by $\tilde{U}_2 =
U_1 \tilde{U}_1$, and so on successively. One can recursively
construct the driving unit cells of time length $2^n T$ as $U_{n+1} =
\tilde{U}_n U_n$. The Fibonacci
sequence~\cite{dumitrescu2018logarithmically,
  maity2019fibonacci,ray2019dynamics} is constructed using the
recursive relation $U_n = U_{n-2} U_{n-1}$ for $n\geq 2$, where the
initial unitary operators are $U_0 = \text{exp}(-i H_+ T)$ and $U_1 =
\text{exp}(-i H_- T)$.  We discuss the case of the Thue-Morse drive in
this section and present the analysis of the Fibonacci drive in
appendix~\ref{app_3}. The results for both cases are similar, but the
dependence of the heating time on the driving frequency is different.

\begin{figure}[h]
  \includegraphics[scale=0.28]{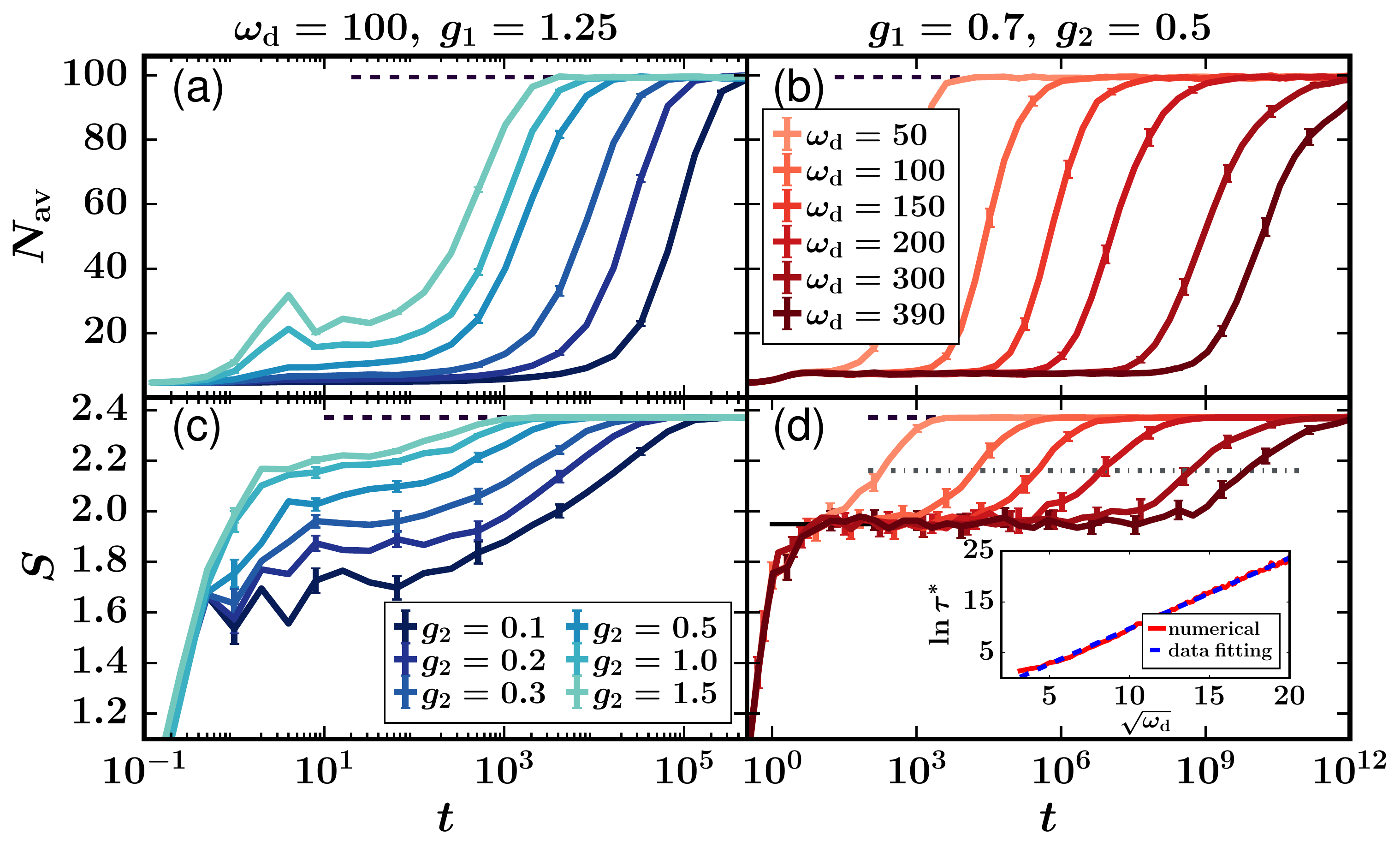}
  \caption{(a)-(b) Average boson number, $N_\text{av}(t)$, and (c)-(d)
    von-Neumann entanglement entropy, $S(t)$, as a function of time
    $t_n=2^n T$ for the anisotropic Dicke model under the Thue-Morse
    quasiperiodic drive. The initial states have low energies, so that
    $\langle E_{\text{in}}\rangle = 3.48 $.  Panels (a,c):
    intermediate driving frequency $\omega_{\text{d}} = 100$, $g_1 = 1.25$,
    and various values of $g_2$. Panels (b,d): $g_1 = 0.7$, $g_2 =
    0.5$ for a chaotic undriven system and various values $\omega_{d}$
    . The inset in Fig.~\ref{fig:TM_adm}(d) shows the scaling of the
    heating time $\tau^{*}$ with $\omega_{\text{d}}$; numerical data are in
    blue and the best fitting, given by $\log\tau^{*}
    =1.4\sqrt{\omega_{\text{d}}} - 4.24$, is in red.  In all panels, the
    driving amplitude is $\Omega=1.0$, $N=10$, and
    $n_{\text{max}}=199$.  The dashed line represents the Page value,
    the black solid line is for the prethermal value, and the
    dashed-dotted line represents the value when the entanglement
    entropy reaches the halfway mark between its prethermal plateau
    and the Page value. }
  \label{fig:TM_adm}
\end{figure}

In Fig.~\ref{fig:TM_adm}, we consider low-energy initial states and
the Thue-Morse driving sequence. We show the dynamics of the average
boson number [Fig.~\ref{fig:TM_adm}(a)-(b)] and the entanglement
entropy [Fig.~\ref{fig:TM_adm}(c)-(d)] for a fixed intermediate value
of the driving frequency $\omega_{\text{d}}$ and various values of the coupling
parameter $g_2$ [Fig.~\ref{fig:TM_adm}(a,c)] and for a fixed $g_2$
associated with the chaotic undriven model and various values of
$\omega_{\text{d}}$ [Fig.~\ref{fig:TM_adm}(b,d)]. All panels exhibit a
prethermal plateau followed by a saturation to the
infinite-temperature state, which contrasts with the results in
Fig.~\ref{fig:Fig03new}~(a,c). The quasiperiodic drive breaks
regularity and induces ergodicity. It causes all cases considered with
intermediate frequency and coupling parameters from the regular to the
chaotic regime to heat up to an infinite temperature.

The prethermal plateau gets longer in time if one increases the
driving frequency or brings the coupling parameters closer to the
regular regime. To quantify the dependence of the prethermal plateau
on the driving frequency, we study the heating time $\tau^{*}$, which
is defined as the time when the entanglement entropy reaches the
halfway mark between its prethermal plateau and the Page
value~\cite{bhakuni2021suppression}, $S(\tau^{*}) \equiv S_p +
[S_{\text{page}} - S_p]/2$. The inset in Fig.~\ref{fig:TM_adm}(d)
shows that for the Thue-Morse drive protocol, the heating time
$\tau^{*}$ grows as a stretched exponential with $\omega_{\text{d}}$, the
best fitting curve corresponding to $\log\tau^{*} =
1.55\sqrt{\omega_{\text{d}}} - 0.695$.  In appendix~\ref{app_3}, we show that
for the Fibonacci drive protocol, the heating time grows exponentially
with the driving frequency as $\log\tau^{*} = 0.125\omega_{\text{d}} - 0.39$.
    
\begin{figure}[h]
  \subfigure{\includegraphics[scale=0.55]{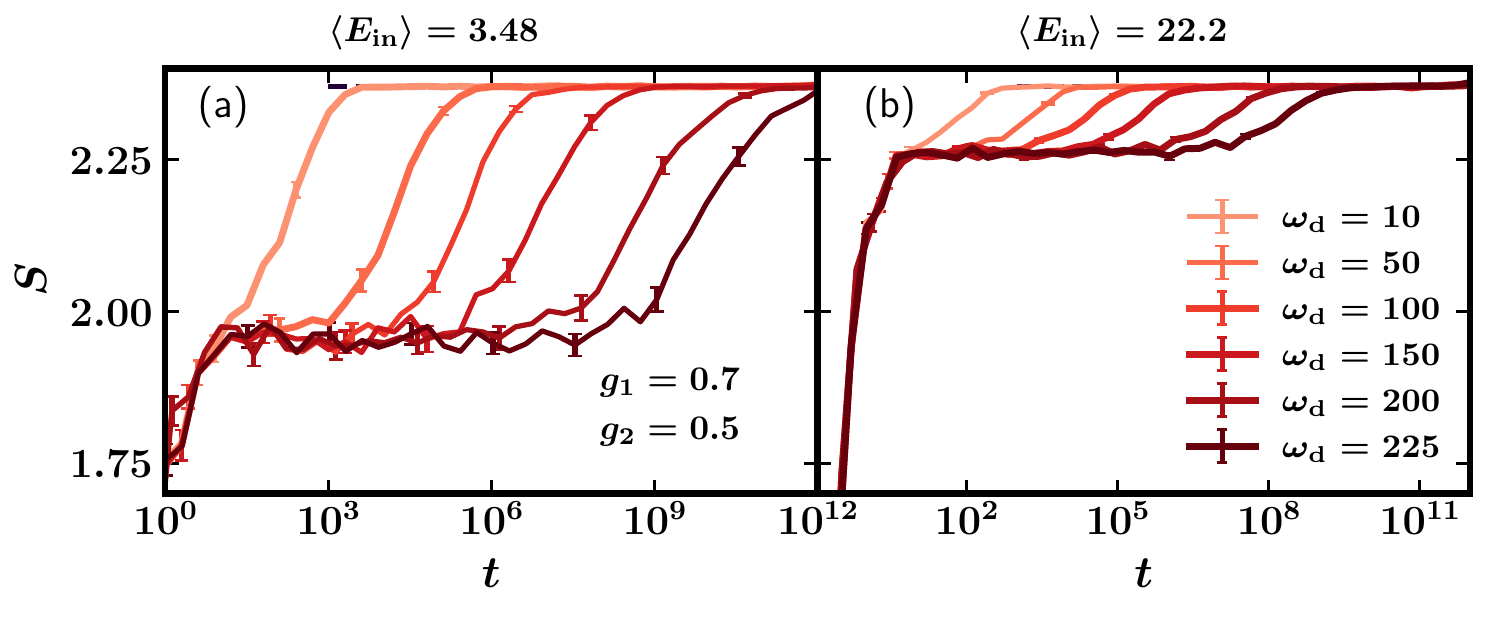}}
  \subfigure{\includegraphics[scale=0.3]{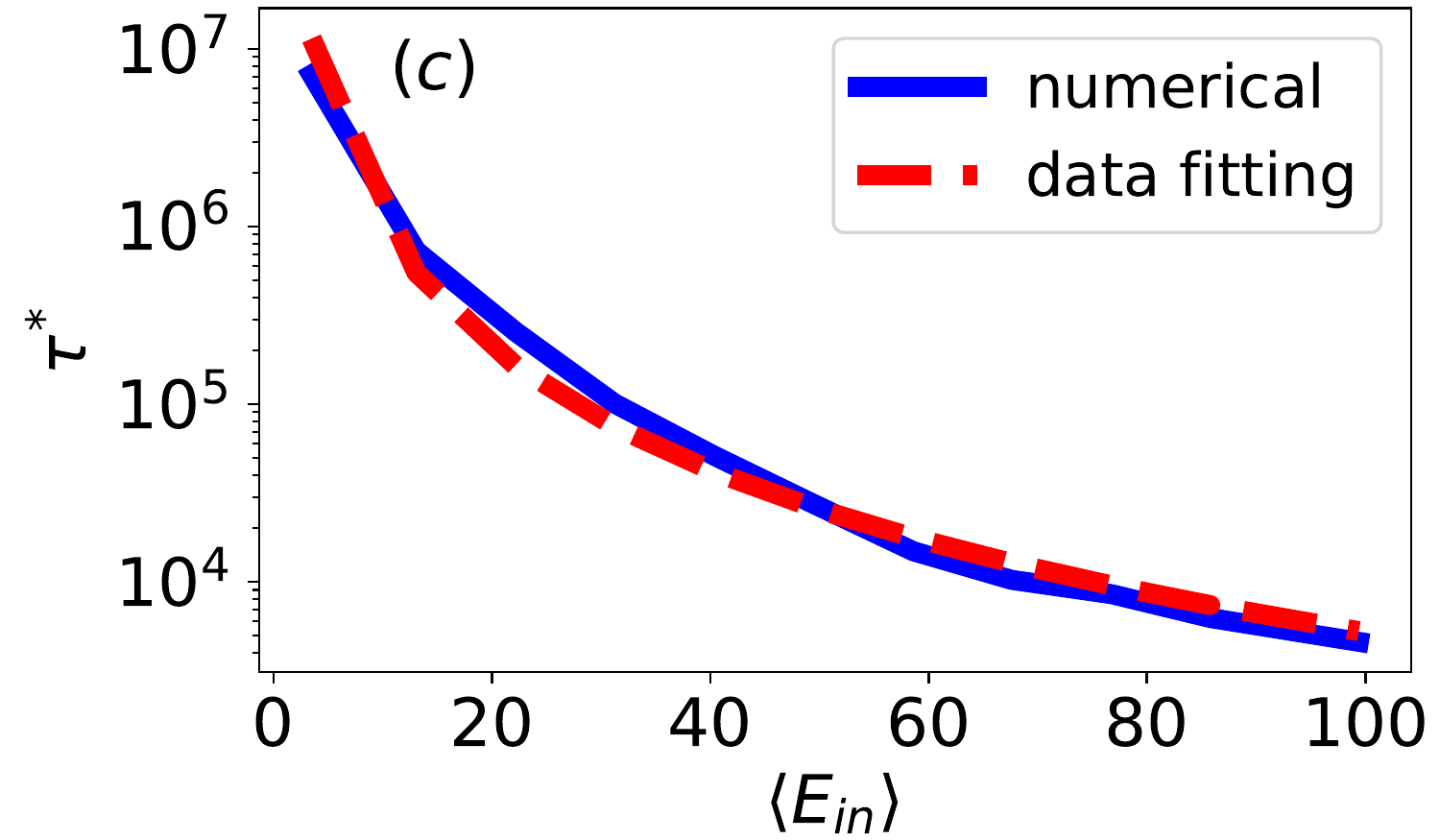}}
  \caption{Entanglement entropy as a function of the sequential time
    $t_n=2^nT$ for the anisotropic Dicke model under the Thue-Morse
    quasiperiodic drive; $g_1=0.7,\ g_2 = 0.5$. Results averaged over
    $50$ initial states with (a) low energy, $\langle
    E_{\text{in}}\rangle= 3.48$, and (b) high energy, $ \langle
    E_{\text{in}}\rangle = 22.2 $. Panel (c): heating time as a
    function of $\langle E_{\text{in}}\rangle$ for a fixed driving
    frequency, $\omega_{\text{d}}=200$; the time scales as $\tau^{*} =
    [1.99\times 10^{8}] {E_{\text{in}}}^{-2.29}$. The driving
    amplitude is $\Omega=1.0$, $N=10$, and $n_{\text{max}}=199$.}
  \label{fig:TM_initial_state}
\end{figure}

In Fig.~\ref{fig:TM_initial_state}, we extend the analysis done in
Fig.~\ref{fig:TM_adm} and investigate how our results are affected by
the rise of the energies of the initial states. In
Figs.~\ref{fig:TM_initial_state} (a)-(b), we plot the evolution of the
entanglement entropy for two sets of initial states with different
energies, respectively given by $\langle E_{\text{in}} \rangle = 3.48$
and $\langle E_{\text{in}} \rangle = 22.2$. As the energy increases,
the prethermal plateau happens at higher values and the heating time
decreases. To check the energy dependence on the heating time, we plot
$\tau^{*}$ as a function of $\langle E_{\text{in}} \rangle$ in
Fig.~\ref{fig:TM_initial_state} (c). We verify that for the Thue-Morse
drive protocol, $\tau^{*}$ decays as $E_{\text{in}}^{-2.29}$. In
appendix~\ref{app_3}, we show that for the Fibonacci drive protocol,
$\tau^{*}$ decays as $E_{\text{in}}^{-4.03}$.

\section{Summary}
\label{sec_4}

We studied the effects that a periodic drive and a quasiperiodic drive
have on the anisotropic Dicke model. While we have verified that some
of the results are similar to those for the driven isotropic Dicke
model (not shown), this work focusses on the more general anisotropic
Dicke model. We list below our four main findings.

(i) Using a periodic drive and the high-frequency Magnus expansion, we
provided a modified condition for the normal to superradiant QPT. By
properly choosing the driving frequency, one can extend the normal
phase.

(ii) We argued that the results for level statistics suggesting
regularity for the periodically driven system under intermediate
frequencies may be an artifact caused by the folding procedure of the
quasienergies back to the principal Floquet zone and the highly
asymmetric shape of the density of states.

(iii) Under the periodic drive, the system saturates to a steady state
value that is not followed by heating to the infinite-temperature
state. The saturation values depend on the energy of the initial
state, the frequency of the drive, and the parameters of the undriven
Hamiltonian. To reach saturation values that indicate near ergodicity,
small driving frequencies are required. Therefore, the non-monotonic
behavior of the saturation values of the average boson number observed
for intermediate driving frequencies imply that these frequencies are
still not small enough to ensure equipartition.

(iv) For the quasiperiodic drives, prethermalization is followed by
heating, ensuring full ergodicity. The heating time $\tau^*$ for the
Fibonacci protocol grows exponentially with the driving frequency
($\tau^* \propto e^{\omega_{\text{d}}})$, while for the Thue-Morse protocol the
growth follows a stretched exponential ($\tau^* \propto
e^{\sqrt{\omega_{\text{d}}}})$. In both cases, the heating time decreases as
the energy of the initial state increases.

Overall, our work shows that the (anisotropic) Dicke model exhibits
properties of genuinely many-body quantum systems that could be
experimentally explored. The absence of heating for the periodic drive
and the long prethermal plateaus for the quasiperiodic drives, for
example, provide scenarios under which non-equilibrium phases of
matter could be hosted.

There are different future directions that we plan to
investigate. Among them, our priorities are the role of dissipation
and a comparison between the quantum and classical dynamics.

\section*{Acknowledgments}
We are grateful to the High Performance Computing(HPC) facility at
IISER Bhopal, where large-scale calculations in this project were
run. P.D. is grateful to IISER Bhopal for the PhD fellowship.
L.F.S. was supported by a grant from the United States National
Science Foundation (NSF, Grant No. DMR-1936006). A.S acknowledges
financial support from SERB via the grant (File Number:
CRG/2019/003447), and from DST via the DST-INSPIRE Faculty Award
[DST/INSPIRE/04/2014/002461].

\appendix
\section{Periodic drive}
\label{app_1}
In this appendix, we derive in Sec.~\ref{app_1a} the analytical
expression of the effective Hamiltonian for the periodically driven
anisotropic Dicke model using the high-frequency Floquet-Magnus
expansion. In Sec.~\ref{app_1b}, we derive the modified equation for
the critical line of the QPT due to the periodic drive.

\subsection{Derivation of the effective Hamiltonian}
\label{app_1a}
We first recall the system Hamiltonian in Eq.~(\ref{eqn:hamiltonian}):
\begin{eqnarray}
	{\cal H}(t) &=& \omega a^{\dagger}a + \omega_{0}J_{z} + \frac{\tilde{g}_{1}(t)}{\sqrt{2j}}(a^{\dagger}J_{-} + a J_{+}) +\nonumber\\
	&& \frac{\tilde{g}_{2}(t)}{\sqrt{2j}}(a^{\dagger}J_{+} + a J_{-}).
\end{eqnarray}
The protocol of the square wave periodic drive applied to the 
system is

\begin{eqnarray}
    \tilde{g}_i(t) &=& g_i + \Omega\hspace{9mm}0< t\leq \frac{T}{2} , \nonumber\\
    \tilde{g}_i(t) &=& g_i - \Omega\hspace{9mm}\frac{T}{2}< t\leq T.\nonumber
\end{eqnarray}
This means that the system is periodically driven by a repeated
two-step sequence that alternates between the time-independent
Hamiltonians $H + V$ and $H - V$ (see main text). The duration of each
step is $T/2$, where $T=2\pi/\omega_{\text{d}}$ is the period of the
driving sequence. The evolution operator at time $t=nT$ is
\begin{align}
U(t=nT) = \left(e^{-iT(H - V)/2}e^{-iT(H + V)/2}\right)^{n}.
\end{align}
Using the Magnus expansion and small $T$, we search for a
time-independent effective Hamiltonian $H_{\text{eff}}$ that
approximately describes the evolution as
\begin{align}
U(t=nT) \approx e^{-inTH_{\text{eff}}}.
\end{align}
Since the driving protocol involves time-independent Hamiltonians, the
Magnus expansion coincides with the Baker-Campbell-Hausdorff
expansion, where the product of two exponentials can be simplified to
\begin{align}
e^X e^Y = e^{\left(X + Y +\frac{1}{2}[X,Y] + \frac{1}{12}[X - Y,[X,Y]] + ...\right)} .
\end{align}
Let
\begin{align}
X = \frac{1}{2}(H - V), \hspace{0.5 cm}
Y = \frac{1}{2}(H + V),
\end{align}
then
\begin{align}
X + Y = H,\hspace{0.5 cm} X - Y = - V,
\end{align}
\begin{align}
[X,Y]=\frac{1}{4}[H - V, H + V]=\frac{1}{2}[H, V], 
\end{align}
and
\begin{eqnarray}
H_{\text{eff}} &=& H + \frac{T}{2i}[X,Y] - \frac{T^{2}}{12}[[X,Y],V] + ...
\end{eqnarray}
After some calculation, we have
\begin{eqnarray}
[X,Y] = \frac{\omega\Omega }{\sqrt{N}}(a^{\dagger}-a)J_x + \frac{\omega_0\Omega}{\sqrt{N}}(a^{\dagger}+a)iJ_y \nonumber\\
 + \frac{(g_1 - g_2)\Omega}{\sqrt{N}}(2iJ_xJ_y - ( a^{\dagger} - a )( a^{\dagger} + a )J_z),
\end{eqnarray}
and
\begin{eqnarray}
[X-Y,[X,Y]] &=& [[X,Y],V] = \nonumber\\ 
&& -\frac{4 \omega\Omega^{2}}{N}J_x^{2} + \frac{2 \omega_0 \Omega^{2}}{N}(a^{\dagger}+a)(a^{\dagger}+a)J_z\nonumber\\
&& + \frac{(g_1-g_2)\Omega^{2}}{N\sqrt{N}}\Big(8(a^{\dagger}+a)J_xJ_z +\nonumber\\
&& (a^{\dagger} - a)({a^{\dagger} + a}^{2})(J_+ - J_-)\Big)  . \nonumber\\
\end{eqnarray}
The first order term, $\frac{T}{2i}[X,Y]=- i\frac{T}{4}[H,V]$, is
imaginary which breaks the time reversal
symmetry~\cite{hetterich2019strong}. Hence we discard the first order
term and consider the second order correction shown above, which leads
to
\begin{eqnarray}
H_{\text{eff}} &=&\omega a^{\dagger}a + \omega_{0}J_{z} + \frac{g_{1}}{\sqrt{N}}(a^{\dagger}J_{-} + a J_{+}) + \frac{g_{2}}{\sqrt{N}}(a^{\dagger}J_{+}\nonumber\\
&& + a J_{-}) - \frac{T^{2}}{12}\Big[-\frac{4 \omega \Omega^{2}}{N}J_x^{2} + \frac{2 \omega_0 \Omega^{2}}{N}(a^{\dagger}+a)\times\nonumber\\
&& (a^{\dagger}+a)J_z + \frac{(g_1-g_2)\Omega^{2}}{N\sqrt{N}}\Big(8(a^{\dagger}+a)J_xJ_z + \nonumber\\ 
&& (a^{\dagger} - a)({a^{\dagger} + a})^{2}(J_+ - J_-)\Big)\Big] .
\label{eqn:adm}
\end{eqnarray}

\subsection{Critical line of the quantum phase transition of the driven system}
\label{app_1b}   

To find the critical line, we first apply the Holstein-Primakoff transformation~\cite{emary2003chaos} to the effective Hamiltonian in Eq.~(\ref{eqn:adm}),
	\begin{equation}
	J_+ = b^{\dagger}\sqrt{2j - b^{\dagger}b},\hspace*{3mm} J_- = \sqrt{2j - b^{\dagger}b}\hspace*{1mm} b,\hspace*{3mm} J_z = b^{\dagger}b - j .
	\label{eqn:HP}
\end{equation}	
In the thermodynamic limit (when the atom number $N\to\infty$), we have
\begin{eqnarray}
H_{\text{eff}}  &=& \omega a^{\dagger}a + \omega_0 b^{\dagger}b + g_1 ( a^{\dagger}b + a b^{\dagger} ) + g_2 ( a^{\dagger}b^{\dagger} + a b )\nonumber\\
&& + \frac{T^2\omega \Omega^2}{12} \left( {b^{\dagger}}^2 + b^2 \right) + \frac{T^2\omega \Omega^2}{6} b^{\dagger} b \nonumber\\
&& + \frac{T^2\omega_0 \Omega^2}{6N} \frac{N}{2} \left( {a^{\dagger}}^2 + a^2 + 2a^{\dagger}a + 1 \right)\nonumber\\
&& - \frac{T^2\omega_0 \Omega^2}{6N} \left( {a^{\dagger}}^2 + a^2  \right) (b^{\dagger} b) - \frac{T^2\omega_0 A^2}{6N} 2 (a^{\dagger} a) (b^{\dagger} b) \nonumber\\
&& - \frac{T^2\omega_0 \Omega^2}{6N} b^{\dagger} b + \frac{T^2 \Delta g \Omega^2}{3 N} \frac{N}{2} ( a + a^{\dagger} ) ( b + b^{\dagger} )\nonumber\\
&& - \frac{T^2 \Delta g \Omega^2}{3 N} ( a + a^{\dagger} ) ( b + b^{\dagger} ) b^{\dagger} b
\end{eqnarray}
We now consider only up to second order terms in the bosonic operators, which means that we neglect the last term of the Hamiltonian in Eq.~(\ref{eqn:adm}). 
Introducing the position and momentum operators for the two bosonic modes as
\begin{eqnarray}
x = \frac{1}{\sqrt{2\omega}}(a^{\dagger} + a),\hspace*{3mm} p_x = i\sqrt{\frac{\omega}{2}}(a^{\dagger} - a),\nonumber\\ y = \frac{1}{\sqrt{2\omega_0}}(b^{\dagger} + b),\hspace*{3mm} p_y = i\sqrt{\frac{\omega_0}{2}}(b^{\dagger} - b),
\end{eqnarray}
we have:
\begin{eqnarray}
H_{\text{eff}}  &=& \left( 1 + \frac{T^2 \Omega^2\omega_0}{6\omega} \right) \frac{1}{2} (\omega^2 x^2 + p_x^2 - \omega)\nonumber\\ 
&& + \left( 1 + \frac{T^2 \Omega^2\omega}{6\omega_0} + \frac{T^2 \Omega^2}{6N} \right) \frac{1}{2} (\omega_0^2 y^2 + p_y^2 - \omega_0)\nonumber\\ 
&& + g_1 \left( \sqrt{\omega\omega_0} x y + \frac{p_x p_y}{\sqrt{\omega\omega_0}} \right) + g_2 \Big( \sqrt{\omega\omega_0} x y \nonumber\\
&& - \frac{p_x p_y}{\sqrt{\omega\omega_0}} \Big) + \frac{T^2 \Omega^2\omega}{12} \left( \omega_0 y^2 - \frac{p_y^2}{\omega_0} \right)\nonumber\\
&& + \frac{T^2 \Omega^2\omega_0}{12}\left( \omega x^2 - \frac{p_x^2}{\omega} \right) + \frac{T^2 \Omega^2\omega_0}{12 N} \left( \omega x^2 - \frac{p_x^2}{\omega} \right)\nonumber\\
&& + \frac{T^2 \Omega^2\omega_0}{12N} \Big( \omega x^2 + \frac{p_x^2}{\omega} + \omega_0 y^2 + \frac{p_y^2}{\omega_0} \Big)\nonumber\\
&& + \frac{T^2 \Omega^2 \Delta g}{3} \sqrt{\omega\omega_0} x y + \frac{T^2 \Omega^2 \Delta g}{3N} \sqrt{\omega\omega_0} x y  .
\end{eqnarray}
To find the critical line for the QPT, we just need to resort to the position part of the equation, which is given by
\begin{eqnarray}
H_{\text{eff}}^{\tilde{x},\tilde{y}}  &=& \frac{1}{2} \left( \tilde{x}^2 + \tilde{y}^2 + \frac{2\gamma}{\alpha\beta\sqrt{\omega\omega_0}} \tilde{x}\tilde{y} \right),
\end{eqnarray}
where, $\tilde{x} = \omega\alpha x$, $\tilde{y} = \omega_0\beta y$, and
\begin{eqnarray}
 \alpha^2 &=& \left( 1 + \frac{T^2 \Omega^2\omega_0}{3\omega} + \frac{T^2 \Omega^2\omega_0}{3N\omega} \right) , \nonumber \\
  \beta^2 &=& \left( 1 + \frac{T^2 \Omega^2\omega}{3\omega_0} \right) , \nonumber \\
   \gamma &=&  \left( g_1 + g_2 + \frac{T^2 \Omega^2 \Delta g}{3} + \frac{T^2 \Omega^2 \Delta g}{3N} \right). \nonumber
   \end{eqnarray}
Introducing normal coordinates,
\begin{equation}
q_+ = \frac{(\tilde{x} + \tilde{y})}{\sqrt{2}},\hspace*{3mm} q_- = \frac{(\tilde{x} - \tilde{y})}{\sqrt{2}},
\end{equation}
we have
\begin{eqnarray}
H_{\text{eff}}^{q_+, q_-} &=& \frac{1}{2} \left[ \left( q_+^2 + q_-^2 \right) + \frac{\gamma}{\alpha \beta \sqrt{\omega\omega_0}} \left( q_+^2 - q_-^2 \right) \right]  \\
&=& \frac{1}{2} \left[ \left( 1 + \frac{\gamma}{\alpha \beta \sqrt{\omega\omega_0}} \right) q_+^2 + \left( 1 - \frac{\gamma}{\alpha \beta \sqrt{\omega\omega_0}} \right) q_-^2 \right] . \nonumber
\end{eqnarray}
From the equation of motion for $q_-$, which is $\ddot{q}_- = - \left( 1 - \frac{\gamma}{\alpha \beta \sqrt{\omega\omega_0}} \right) q_-$, one gets the equation of the critical line for the QPT in  the $g_1-g_2$ plane, 
\begin{equation}
1 - \frac{\gamma}{\alpha \beta \sqrt{\omega\omega_0}} = 0 .
\end{equation}
Introducing the notation $\delta=\frac{T^2 \Omega^2}{3}$, we have
\begin{eqnarray}
g_1 + g_2 &=& \Big( 1 + \frac{\delta\omega}{2\omega_0} + \frac{\delta\omega_0}{2\omega} + \frac{\delta\omega_0}{2N\omega} \Big)\sqrt{\omega\omega_0}\nonumber \\
&& - \delta\Delta g - \frac{\delta\Delta g}{N},
\end{eqnarray}
where $\Delta g = g_1-g_2$ and we have not considered the other higher order terms as $\delta<<1$.
In the thermodynamic limit ($N\to \infty$), we finally obtain
\begin{eqnarray}
g_1 + g_2 &=& \Big( 1 + \frac{\delta\omega}{2\omega_0} + \frac{\delta\omega_0}{2\omega} \Big)\sqrt{\omega\omega_0} - \delta\Delta g.
\end{eqnarray}
or,
\begin{eqnarray}
(1+\delta)g_1 + (1-\delta)g_2 &=& \Big( 1 + \frac{\delta}{2}\Big( \frac{\omega}{\omega_0} + \frac{\omega_0}{\omega} \Big) \Big)\sqrt{\omega\omega_0}\nonumber \\
\end{eqnarray}
and hence,
\begin{eqnarray}
g_2 &=& \left( \frac{1+\tilde{\delta}}{1-\delta} \right)\sqrt{\omega\omega_0} - \left( \frac{1+\delta}{1-\delta} \right) g_1,
\end{eqnarray}
where $\tilde{\delta} = \frac{\delta}{2}\left( \frac{\omega}{\omega_0} + \frac{\omega_0}{\omega} \right)$.
or,
\begin{eqnarray}
g_2 &=& \tilde{\chi}\sqrt{\omega\omega_0} - \chi g_1,
\end{eqnarray}
where, $\chi = \left( \frac{1+\delta}{1-\delta} \right)$ and $\tilde{\chi} = \left( \frac{1+\tilde{\delta}}{1-\delta} \right)$.

\section{Periodic drive}
\label{app_2}

This appendix extends the results presented in Fig.~\ref{fig:Fig03new}
of the main text.

\subsection{Dependence on the coupling parameters}
\begin{figure}[h]
  \includegraphics[scale=0.28]{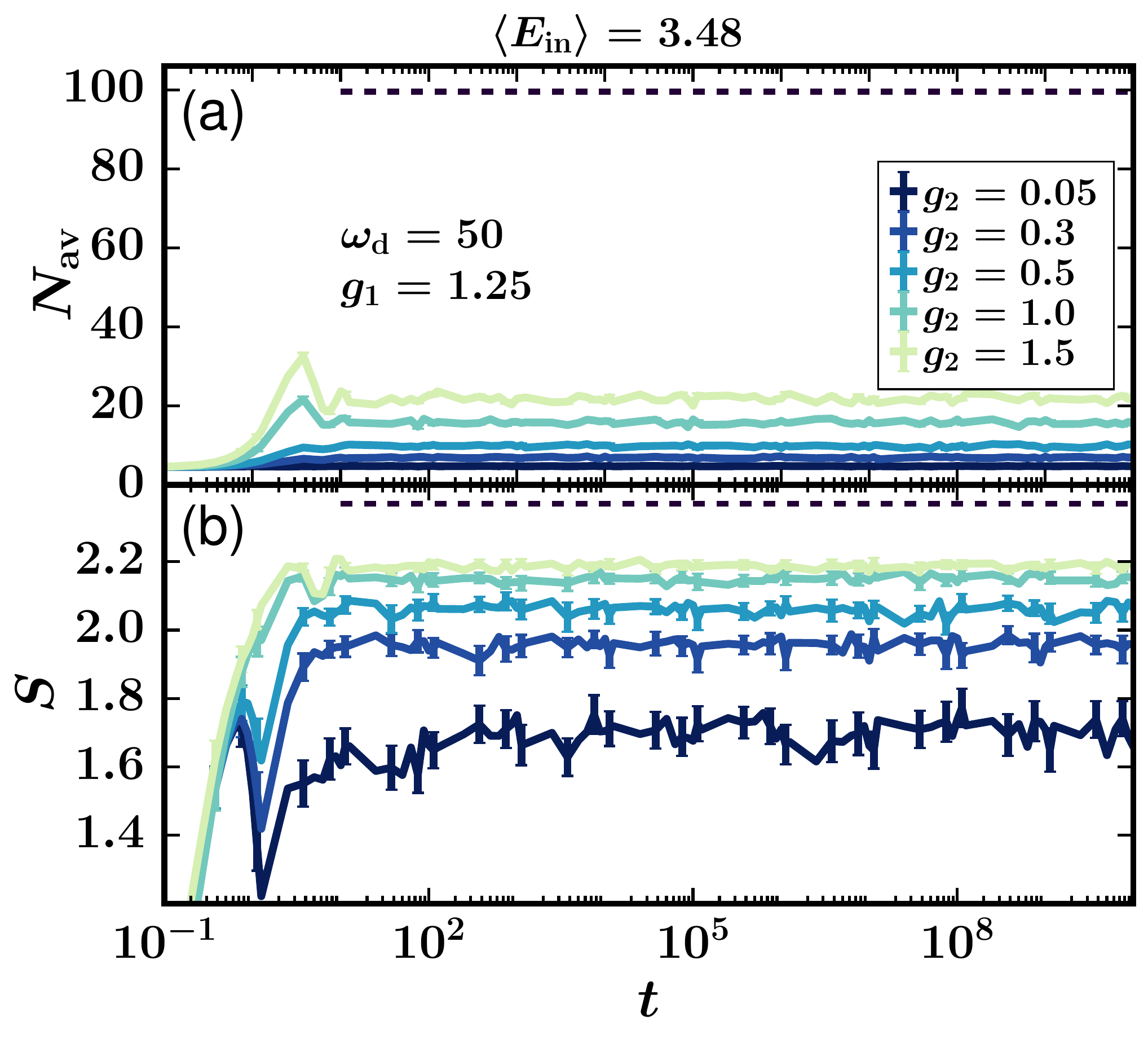}
  \caption{ (a) Average boson number, $N_\text{av}(t)$, and (b) von-Neumann entanglement entropy $S(t)$ as a function of the stroboscopic time $t_n=n T$ for the periodically driven anisotropic Dicke model. The initial states have low energies, so that $\langle E_{\text{in}}\rangle = 3.48$, as in Figs.~\ref{fig:Fig03new}(a,b). The driving frequency is fixed at an intermediate value, $\omega_{\text{d}}=50$ and various values of $g_2$ are considered, as indicated. The black dashed line indicates the results for the infinite-temperature state. 
    For both panels, $g_1 = 1.25$, the driving amplitude is $\Omega=1.0$, $N=10$, and $n_{\text{max}}=199$. }
  \label{fig:FigApp2}
\end{figure}

To complement Figs.~\ref{fig:Fig03new}(a,c) of the main text and
support the discussion made there about the dependence of the
saturation values of $N_\text{av} (t)$ and $S(t)$ on the coupling
parameters, we show in Fig.~\ref{fig:FigApp2} the evolution of the
average boson number [Fig.~\ref{fig:FigApp2}(a)] and the entanglement
entropy [Fig.~\ref{fig:FigApp2}(b)] for low-energy initial states and
a fixed intermediate value of the driving frequency
$\omega_{\text{d}}$. Various values of the coupling parameter $g_2$
are considered, so that the undriven Hamiltonian goes from the regular
to the chaotic regime.

As explained in the main text, for an intermediate frequency and
low-energy initial states, the periodic drive is unable to bring
$N_\text{av} (t)$ and $S(t)$ close to the results of the
infinite-temperature state, at least not for the very long times that
we studied. The saturation values of the two quantities are always
below $N^{\infty}$ and $S_{\text{Page}}$ and, as shown in
Fig.~\ref{fig:FigApp2}(a,c), they further decrease, as we decrease
$g_2$ and the undriven model is brought closer to the regular regime.

\subsection{Dependence on the initial state energy}
\label{app_2b} 

\begin{figure}[h]
  \includegraphics[scale=0.28]{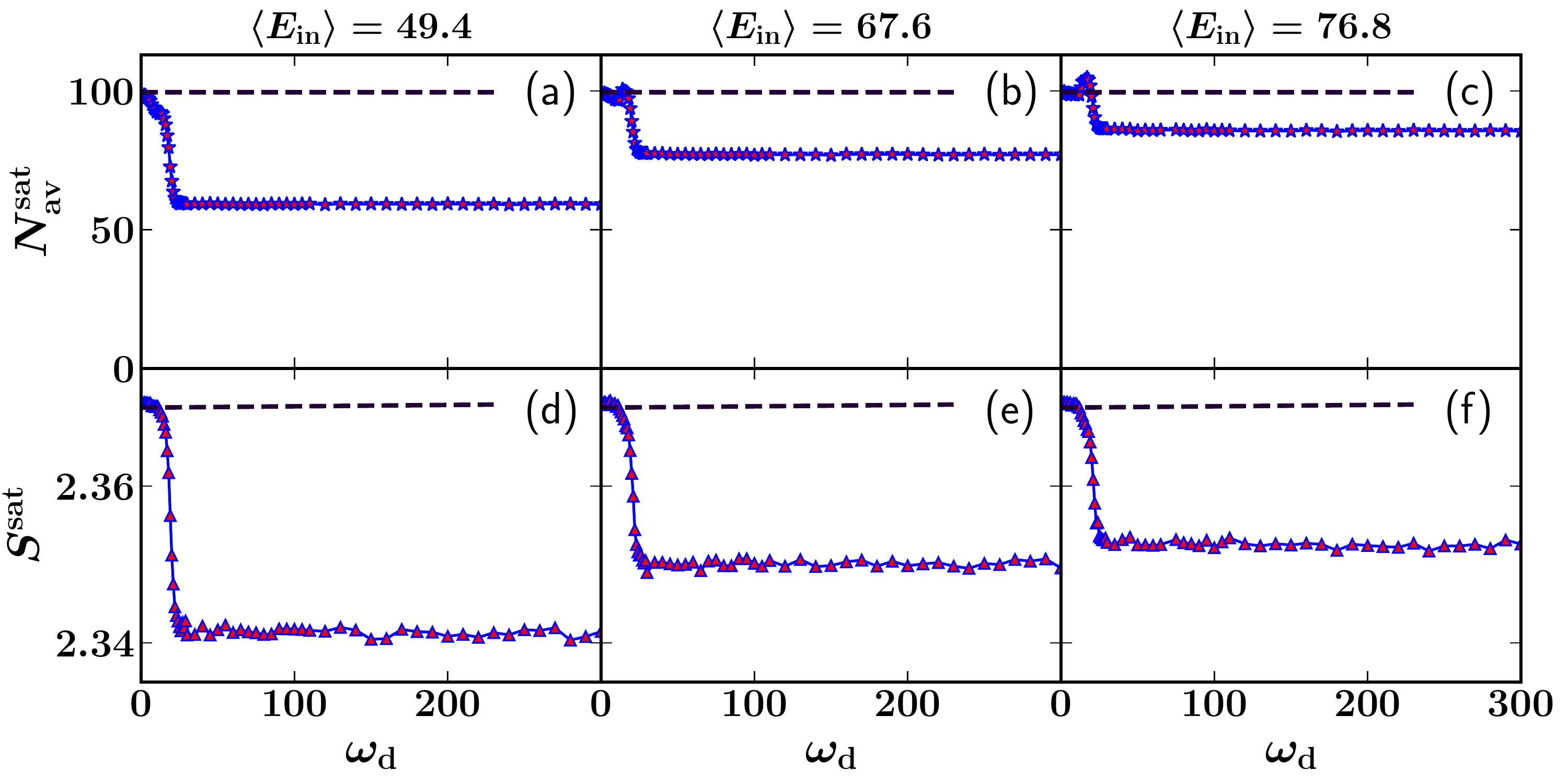}
  \caption{The saturation value of the average boson number (top panels) and of the von-Neumann entanglement entropy between spins and bosons (bottom panels) as a function of the driving frequency. The black dashed line indicates the infinite-temperature result. We fix $g_1=1.25,\ g_2 = 1.0$, $\Omega=1$, and atom number $N=10$, bosonic cut-off $n_{\text{max}}=199$. }
  \label{fig:Nav_wd}
\end{figure}

Figure~\ref{fig:Nav_wd} shows the saturation values of average boson
number (top panels: Fig.~\ref{fig:Nav_wd}(a)-(c)) and of the
von-Neumann entanglement entropy (bottom panels:
Fig.~\ref{fig:Nav_wd}(d)-(f)) as a function of the driving frequency
for different values of the initial state energy. While
$S^{\text{sat}}$ grows monotonically towards the infinite-temperature
result as $\omega_{\text{d}}$ decreases, the same does not happen for
$N_{\text{av}}^{\text{sat}}$ when the energy of the initial state is
high. There is a very narrow range of driving frequencies where
$N_{\text{av}}^{\text{sat}}> N_{\text{av}}^{\infty}$. This implies
that the value of the driving frequency is still not small enough to
ensure equipartition. As $\omega_{\text{d}}$ decreases from $\infty$, the fact
that $N_{\text{av}}^{\text{sat}}$ crosses $N_{\text{av}}^{\infty}$,
before becoming larger than it, is not caused by ergodicity, but by
the significant number of states contributing to the dynamics, which
have average boson number in the vicinity of $N_{\text{av}}^{\infty}$.

\section{Fibonacci sequence}
\label{app_3}

\begin{figure}[h]
    \includegraphics[scale=0.28]{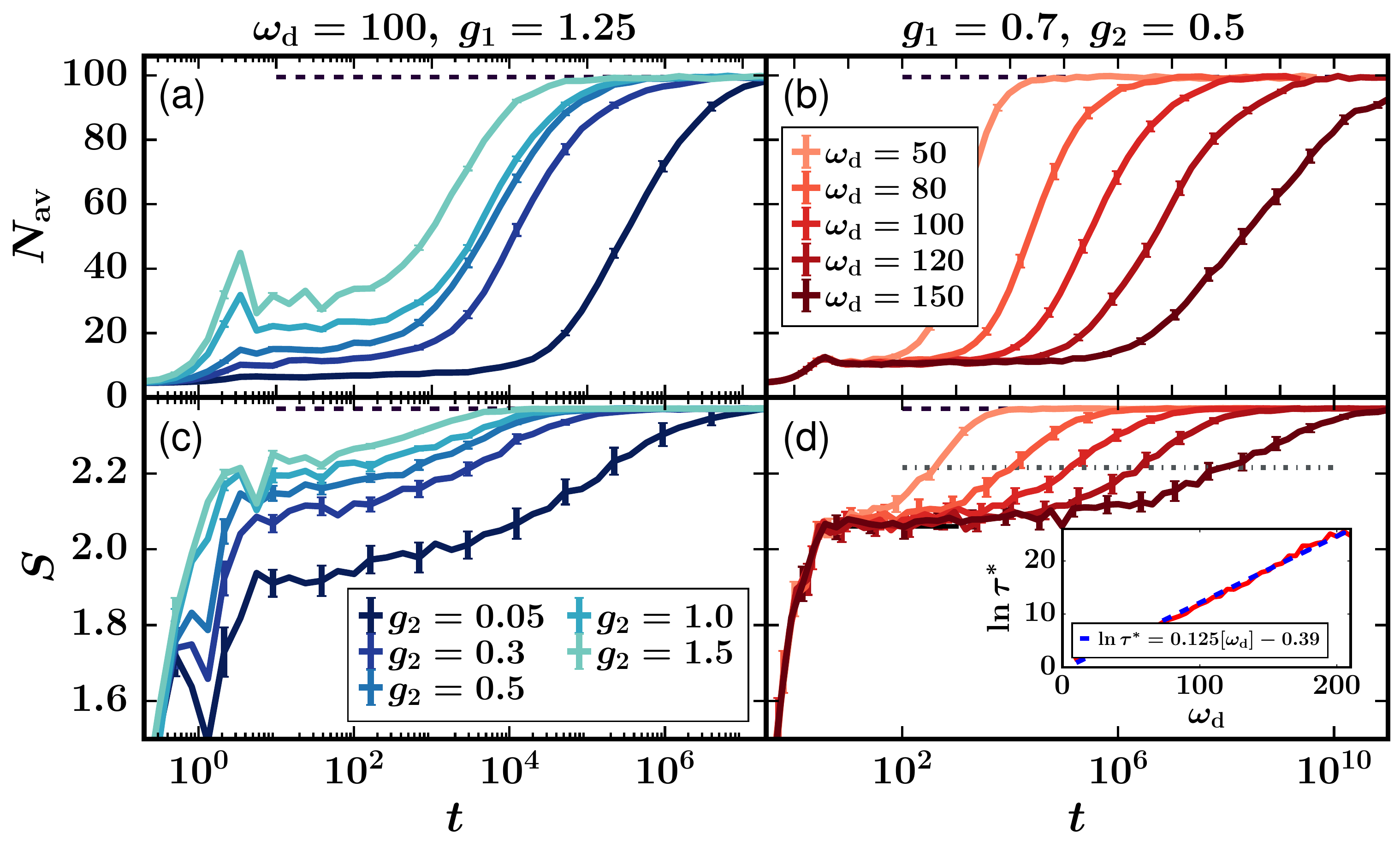}
     \caption{(a)-(b) Average boson number $N_\text{av}(t)$ and (c)-(d)
    von-Neumann entanglement entropy $S(t)$ as a function of time $t_n=t_{n-1}+t_{n-2}$ for the anisotropic Dicke model under the Fibonacci quasiperiodic drive. The initial states have low energies, so that $\langle E_{\text{in}}\rangle = 3.48$. Panels (a,c): intermediate driving frequency $\omega_{\text{d}} = 100$, $g_1 = 1.25$, and various values of $g_2$. Panels (b,d): $g_1 = 0.7$, $g_2 = 0.5$ for a chaotic undriven system and various values
    $\omega_{\text{d}}$ . The inset in panel (d) shows the scaling of
    the heating time $\tau^{*}$ 
    with $\omega_{\text{d}}$; numerical data are in blue and the best fitting, given by $\log\tau^{*} =
    0.125\omega_{\text{d}} - 0.39$, is in red. In all panels, the
    driving amplitude is $\Omega=1$, $N=10$, and
    $n_{\text{max}}=199$. 
    In this figure the dashed line represent the page value, the black solid line is for the prethermal value whereas the dashed-dotted line represents the value when the entanglement entropy reaches the halfway mark between its prethermal plateau and the Page value.
    }
\label{fig:FB}
\end{figure}

\begin{figure}[h]
  \subfigure{\includegraphics[width=0.45\textwidth]{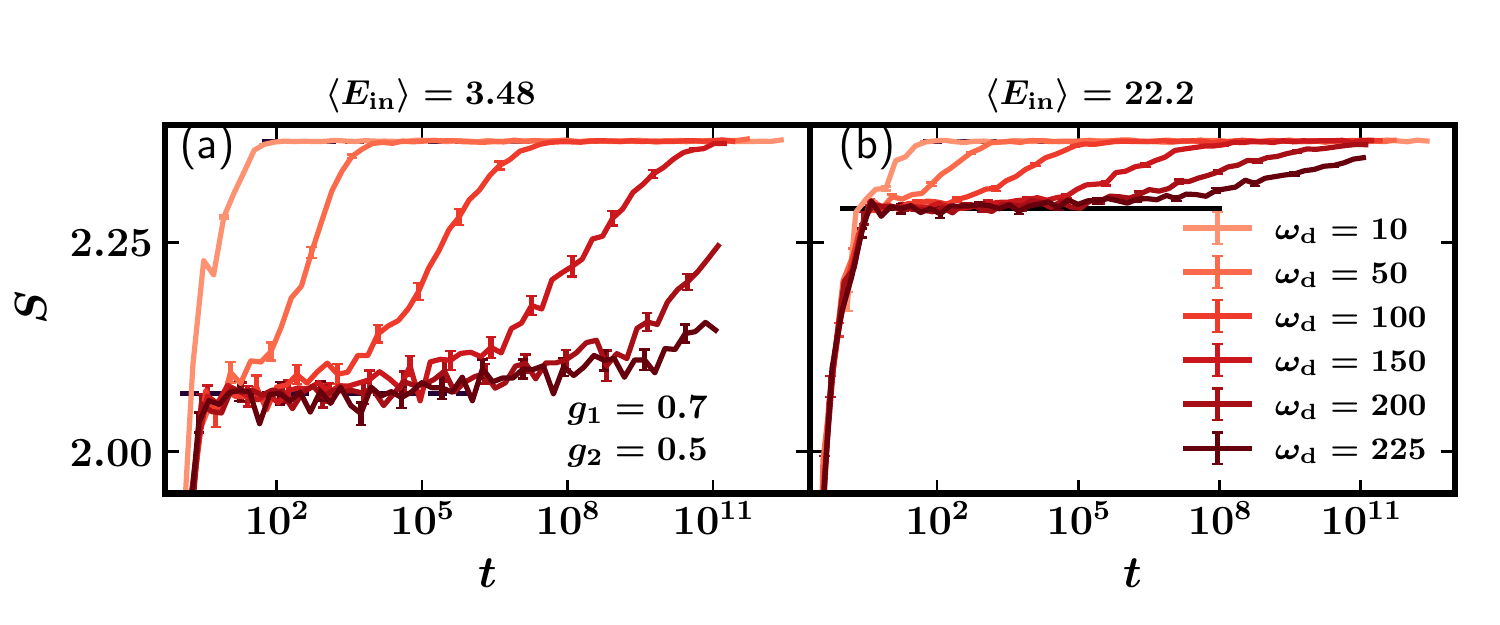}\label{fig:num_gs}}
  \subfigure{\includegraphics[width=0.25\textwidth]{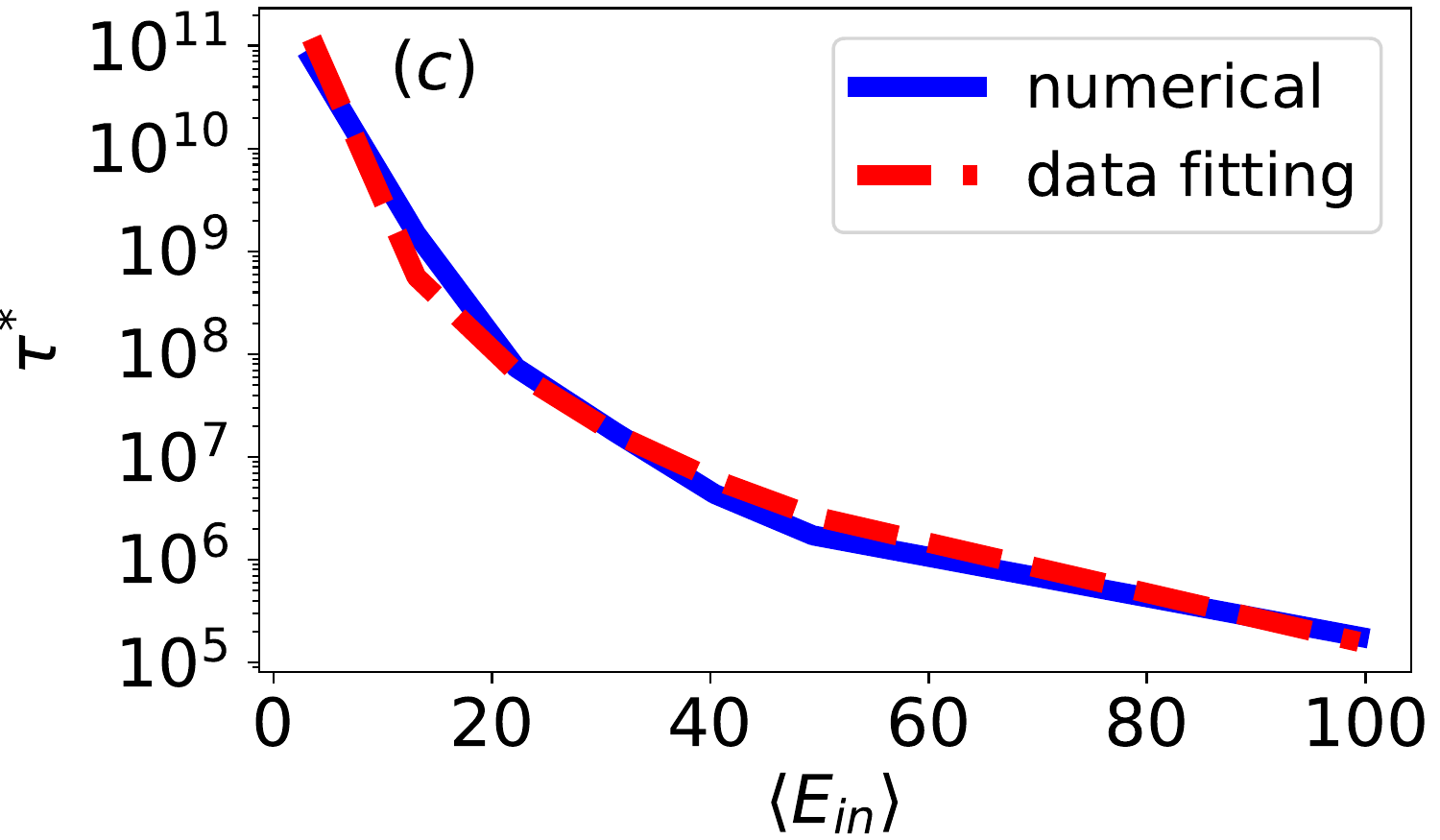}\label{fig:num_gs}}
  \caption{Entanglement entropy  as a function of sequential time $t_n=t_{n-2} + t_{n-1}$ for the anisotropic Dicke model under the Fibonacci quasiperiodic drive;  $g_1=0.7,\ g_2 = 0.5$. Results averaged over $50$ initial states with (a) low energy, $\langle E_{\text{in}}\rangle= 3.48$ and (b) high energy, $ \langle E_{\text{in}}\rangle = 22.2 $.
 Panel (c): heating energy as a function of $\langle E_{\text{in}}\rangle$ for a fixed driving frequency $\omega_{\text{d}}=200$ and it scales as: $\tau^{*} = [1.8034\times 10^{13}] {E_{\text{in}}}^{-4.03}$. The driving amplitude is $\Omega=1.0$, $N=10$ and  $n_{\text{max}}=199$.}
      \label{fig:FB_initial_states}
    \end{figure}

The results shown here for the Fibonacci quasiperiodic drive are
similar to those shown in Sec.~\ref{sec_3APERIODIC} for the Thue-Morse
quasiperiodic drive, with the difference that there $\tau^{*} \propto
\exp(\sqrt{\omega_{\text{d}}})$, while here $\tau^{*} \propto
\exp(\omega_{\text{d}})$. Figure~\ref{fig:FB} is equivalent to
Fig.~\ref{fig:TM_adm}, and Fig.~\ref{fig:FB_initial_states} is
equivalent to Fig.~\ref{fig:TM_initial_state}.

In Fig.~\ref{fig:FB}, we consider low-energy initial states and the
Fibonacci driving sequence. We show the dynamics of the average boson
number [Fig.~\ref{fig:FB}(a)-(b)] and the entanglement entropy
[Fig.~\ref{fig:FB}(c)-(d)] for a fixed intermediate value of the
driving frequency $\omega_{\text{d}}$ and various values of coupling parameter
$g_2$ [Fig.~\ref{fig:FB}(a,c)] and for a fixed $g_2$ associated with
the chaotic undriven model and various values of $\omega_{\text{d}}$
[Fig.~\ref{fig:FB}(b,d)]. All panels exhibit a prethermal plateau
followed by the saturation to the infinite-temperature state.  The
prethermal plateau gets longer in time as we increase the driving
frequency or bring the coupling parameters closer to the regular
regime. The anisotropic Dicke model under this quasiperiodic drive
heats up exponentially slowly, as shown in the inset of
Fig.~\ref{fig:TM_adm}(d), where the heating time grows with
$\omega_{\text{d}}$ as $\log\tau^{*} = 0.125\omega_{\text{d}} - 0.39$.
    
In Fig.~\ref{fig:FB_initial_states}(a)-(b), we compare the evolution
of the entanglement entropy for two different initial states energies,
respectively $\langle E_{\text{in}} \rangle = 3.48$ and $\langle
E_{\text{in}} \rangle = 22.2$. As the energy increases, the prethermal
plateau happens at higher values and the heating time decreases. To
check the energy dependence on the heating time, we plot $\tau^{*}$ as
a function of $\langle E_{\text{in}} \rangle$ in
Fig.~\ref{fig:FB_initial_states} (c) and we verify that $\tau^{*}$
decays as $E_{\text{in}}^{-4.03}$.

\bibliography{ref_new}
\clearpage

\end{document}